\documentclass[12pt,a4paper]{article}
\usepackage{amssymb}
\usepackage{amsmath}
\usepackage{mathrsfs} 
\usepackage[super,compress]{cite}
\usepackage{graphicx} 
\usepackage{anyfontsize} 
\usepackage[font=footnotesize,labelfont=it,labelsep=period]{caption} 
\usepackage{float} 

\input{tcilatex}

\begin{document}

\begin{center}
\textbf{Study of the Quartic Anharmonic Oscillator Using the  
   System's Wave Function Expansion in the Oscillator Basis}{\tiny \bigskip }

\textbf{V. A. Babenko\footnote{%
E-mail: pet2@ukr.net} and A. V. Nesterov}

\textit{Bogolyubov Institute for Theoretical Physics, NAS of Ukraine, Kyiv}
\end{center}

\thispagestyle{empty}

\noindent The quantum quartic anharmonic oscillator with the Hamiltonian $H=\frac{1%
}{2}\left( p^{2}+x^{2}\right) +\lambda x^{4}$ is a classical and fundamental 
model that plays a key role in various branches of physics, including 
quantum mechanics, quantum field theory, high-energy particle physics, and other areas.
To study this model, we apply a method based on
 a convergent expansion of the system's wave function in a complete set of harmonic
oscillator eigenfunctions --- namely, the basis of eigenfunctions $%
\varphi^{(0)}_n$ of the unperturbed Hamiltonian $H^{(0)}=\frac{1}{2}\left(
p^{2}+x^{2}\right)$. This approach enables a thorough analysis and
calculation of the oscillator's physical characteristics. We demonstrate  
very good convergence of all calculated quantities with respect to the
number of basis functions included in the expansion, over a wide range of
  $\lambda$ values. We have computed the energies of the
ground and the first six excited states for a broad range of the coupling
constant $\lambda$, and also calculated and constructed the corresponding 
wave functions. Additionally, we propose and detail
an improved version of the expansion method using a modified optimized 
oscillator basis with variable frequency. This modification significantly
accelerates the convergence of expansions across the entire range of $%
\lambda $, thereby greatly enhancing the efficiency of the
 method and allowing accurate calculations with a very small number of
expansion functions $N\lesssim 10$. As a result, 
this modified approach provides an
essentially  complete, simple, and efficient solution to the problem of the
anharmonic oscillator, enabling straightforward computation of all its
physical properties --- including the energies and wave functions of 
both ground and excited states --- for 
arbitrary values of $\lambda$.

\vspace*{1.1em}

\vspace*{1.1em}

\begin{center} 
{\textbf{1. Introduction}} 
\end{center}  
    The quantum anharmonic oscillator is a classic, traditional, and widely used
model that finds many theoretical and practical applications in various
fields of quantum physics [1--10]. Throughout
the development of quantum theory, this model has generated
immense interest and has been the subject of numerous studies due to its
diverse important properties and characteristics, as well as its potential
applications in describing a large number of different quantum systems and
phenomena. This is a result of the significant fact that the  
anharmonic oscillator model is a generalization and extension of the
harmonic oscillator model and can describe processes and their
characteristics in many quantum systems with vibrational degrees of freedom.
Moreover, these vibrational degrees of freedom are, evidently, found in a
multitude of systems. As a consequence, its practical applications 
 span broad areas of quantum physics and attract significant interest [1--10]. 
Among the model's many applications,
notable ones include the description of molecular and atomic vibrations in
quantum chemistry and atomic-molecular physics, crystal lattice vibrations
in solid-state theory, certain diffusion processes, and applications in
laser theory. It is worth emphasizing that the anharmonic oscillator 
is also a foundational model used to evaluate different quantum-mechanical 
approximation and perturbation methods. In addition, this model can serve 
as an approximation for more complex potentials.    
     For several reasons, the anharmonic oscillator is of particular
interest and importance for quantum field theory, particle physics, and
nuclear theory [1--15]. 
According to many authors, this model is one of the
simplest yet quite realistic models of quantum field theory, 
possessing many 
characteristic quantum field features. Therefore, it can
serve as an important example and illustration for studying and applying 
 various quantum field methods, effectively modeling complex fields in 
one-dimensional space-time. 
More realistic and multidimensional quantum-mechanical 
and quantum-field models are expected to share 
the same fundamental properties as 
 the anharmonic oscillator,
thereby making it a very convenient model for studying these properties. In
particular, some of the most important issues to be explored using the
anharmonic oscillator include fundamental problems such as the
divergence of 
perturbation series and
the challenge of describing strong coupling effects,
i.e., describing system properties in the non-perturbative region at high
values of the coupling constant. 
    It is also noteworthy that, in nuclear theory, certain collective models 
under specific simplifications lead to Hamiltonians of the anharmonic 
oscillator type [11--15]. This naturally reflects the 
aforementioned ability of the anharmonic 
oscillator to describe vibrational degrees of freedom in 
various quantum systems, including surface vibrations 
and nucleonic oscillations in atomic nuclei. 

   Thus, the anharmonic oscillator model has been the subject of numerous
intensive studies throughout the development of quantum physics.
Accordingly, hundreds  of papers have been dedicated to its study ---
see, for example, earlier publications [1, 2, 16--29], 
a series of more recent works [9, 11--15, 30--44], and a
newly published extensive monograph [10] devoted to this topic.  
Consequently, the anharmonic oscillator
consistently attracts significant research interest due to its
relatively simple yet substantially nontrivial and demonstrative nature, as
well as its potential practical applications. Therefore, it is practically
impossible to present a complete list of papers dedicated to this topic.
Nevertheless, some bibliographies on the anharmonic oscillator, which
include references to several recent studies, are available in some  
papers [9, 38] and in a just-published monograph [10] on this subject. 
     Significant further interest in the anharmonic oscillator and the growing
relevance of this issue arose following the publication of the
 fundamental paper by Bender and Wu [1]. In this study,
the divergence of the perturbation theory series for the ground-state energy 
$E_{0}\left( \lambda \right) $ of the quartic anharmonic oscillator was
rigorously proved across the entire complex plane of the coupling constant 
$\lambda $, except at the origin. Since similar types of
divergence were subsequently also found and studied in other
quantum-mechanical and quantum-field systems, the work of Bender and Wu
essentially led to the emergence of an entirely new research direction,
commonly known as Large-Order Perturbation
Theory [4, 6--10]. In fact, Bender and Wu
were the first to rigorously prove the divergence of the perturbation theory
series for a specific model of quantum field theory, albeit one of the
simplest. Notably, the quantum-field nature of the anharmonic oscillator
model gives rise to two important, yet essentially opposite, possibilities 
for research and application, as implemented by Bender and Wu 
[1, 18, 19]. On the one hand, studying this model using 
classical mathematical methods of quantum mechanics and differential
equations --- such as the WKB approximation --- enables the establishment and
verification of various properties applicable to it as a model of quantum field
theory and, more broadly, for its potential generalizations. On the other hand,
 the model's properties can be explored using novel
methods developed specifically in quantum field theory, as demonstrated by
Bender and Wu. 
Thus, all the aforementioned features make the model particularly 
useful for exploring quantum field-theoretic and phenomenological 
aspects of strong interaction physics.

   In our previous works [45, 46], the quartic anharmonic oscillator 
was studied based on summing the divergent Rayleigh-Schr\"{o}dinger
perturbation series using our proposed improved method of Pad\'{e}
approximants with their averaging. This approach not only offered multiple
advantages but also enabled, for the first time, the correct asymptotic
behavior of the energy levels at infinity as the oscillator's coupling
constant $\lambda $ increases. In the present work, we conduct a 
 comprehensive analysis of the quartic anharmonic oscillator, 
including detailed calculations of its properties and
characteristics, by using the convergent expansion of the system's wave function  
 in the complete set of harmonic
oscillator eigenfunctions, namely the eigenfunctions of the unperturbed
Hamiltonian $H\left( \lambda =0\right) $. This method of expansion in the
oscillator basis has been successfully and repeatedly applied 
 by us within the framework of the 
 Resonating Group Method to study the 
properties of light atomic nuclei [47--50]. Overall, the method of expansion in
the oscillator basis is an established traditional approach to solving many
quantum-mechanical problems, which has been repeatedly and successfully used
in numerous theoretical tasks. See, for example, the extensive monograph 
[51] on this topic and the numerous examples and references therein.
     However, to the best of our knowledge, a consistent and 
clear application of 
this technique to the anharmonic oscillator problem
is currently lacking in the
literature, although procedures involving diagonalization of the
Hamiltonian and the variational methods of several studies [41, 52, 53]
are quite close to this approach. It is worth noting that the very recent study 
[41], while closely related to the application of the given method, is
actually focused on a double-well potential rather than the anharmonic
oscillator. Since, in the context of the quartic oscillator, this study [41] 
presents only one calculation at an extremely low coupling constant value, 
$\lambda =1/4$, within the ultra-weak coupling range, it lacks
exploration and calculations for other values of the coupling constant
applicable to the anharmonic oscillator. Therefore, it does not offer  
 a comprehensive treatment of the problem. 
Moreover, in the present work,
within the framework of the oscillator basis expansion approach, we also
 thoroughly examine the important issue of significantly
improving the description of the system's calculated properties and characteristics 
  in the non-perturbative region of strong and superstrong
coupling. This enhancement is achieved by means of a newly proposed method
that accelerates the convergence of expansions in the oscillator basis. 
\vspace*{0.5em}

\begin{center}
{\textbf{ 2. Foundations of the Theory: Quartic Anharmonic Oscillator and Oscillator
Basis Expansion }}  
\end{center}
The model of the 1D quartic anharmonic quantum oscillator with
the Hamiltonian: 
\begin{equation}
H=H^{\left( 0\right) }+U=H^{\left( 0\right) }+\lambda V=\frac{1}{2}\left(
p^{2}+x^{2}\right) +\lambda x^{4}  \tag{1}
\end{equation}%
is, as previously noted, one of the classic traditional models of quantum
mechanics and quantum field theory [1--10]. The
perturbation operator $U\left( x\right) =\lambda V\left( x\right) =\lambda
x^{4}$ in this model is directly proportional to the fourth power of the
coordinate $x$. At the initial stage of the study, we will 
 follow common conventions and  
 use a system of units in
which the fundamental parameters of the oscillator (its mass, frequency, and
reduced Planck constant) are set equal to one: $m=1$, $\omega =1$, $\hbar =1$%
. The stationary Schr\"{o}dinger equation in the coordinate representation
for the system's wave function $\psi \left( x\right) $ in this case has the
usual form: 
\begin{equation}
H\psi =E\psi ~,  \tag{2}
\end{equation}%
where the squared momentum operator $p^{2}$ can be expressed through the
coordinate $x$, considering the adopted conventions, as $p^{2}=-\frac{d^{2}}{%
dx^{2}}$. An important characteristic parameter $\lambda $, which describes
the deviation of the system's total Hamiltonian (1) from the Hamiltonian of
the well-known standard harmonic oscillator $H^{\left( 0\right) }=\frac{1}{2}%
\left( p^{2}+x^{2}\right) $, is usually referred to as the coupling constant
or the anharmonicity parameter. 
     In the initial phase of the research, it is
natural to consider the coupling constant $\lambda $ as a small, real, and
positive parameter. However, the description and calculation of the system's
characteristics at large real values of the parameter $\lambda $, i.e., in
the region of strong coupling, also represent significant fundamental and
practical interest. To fully establish and theoretically investigate all
the essential properties of the system, it is necessary to study the
behavior of its characteristics, such as energy $E$ and wave function $\psi $%
, across the entire complex plane of the parameter $\lambda $ variation 
[1, 10, 17]. Note also that the wave function $\psi \left( x\right) $, as
seen from (1), (2), depends on the system's energy $E$ and the coupling
constant $\lambda $ as parameters --- $\psi \left( x\right) =\psi _{E}\left(
\lambda ;x\right) $.

    Equation (2) in the coordinate representation is a second-order ordinary
differential equation for the system's wave function $\psi \left( x\right) $
and, as is well known, has physical quadratically integrable solutions,
called eigenfunctions, only at certain discrete energy values $%
E_{n}=E_{n}\left( \lambda \right) $, forming a spectrum of eigenvalues $%
\left\{ E_{n}\left( \lambda \right) \right\} _{n=0}^{\infty }$. The general
solution of the second-order differential equation (2) can be represented as
a linear combination, $\psi \left( x\right) =C_{1}\psi _{1}\left( x\right)
+C_{2}\psi _{2}\left( x\right) $, of two linearly independent solutions, $%
\psi _{1}\left( x\right) $ and $\psi _{2}\left( x\right) $, with certain
constants $C_{1}$ and $C_{2}$, where one solution decreases at infinity as $%
x\rightarrow +\infty $, while the other solution increases. Based on
physical considerations, the increasing solution (for definiteness, for
example, $\psi _{2}\left( x\right) $) should be discarded by setting one of
the constants $C_{i}$ in the linear combination of solutions to zero (in
this case, $C_{2}=0$). The remaining arbitrary coefficient $C_{1}$ is
theoretically determined by the condition of normalizing the wave function $%
\psi \left( x\right) $ of the system's bound state to unity: 
\begin{equation}
\tint\limits_{-\infty }^{\infty }\psi ^{2}\left( x\right) dx=1~.  \tag{3}
\end{equation}%
In this case, the wave function  is chosen to be real
and quadratically integrable, and its asymptotic behavior
at infinity for the selected decreasing solution, according to the
differential equation (2), as demonstrated in [17, 54], takes the form: 
\begin{equation}
\psi \left( x\right) \underset{x\rightarrow +\infty }{\simeq }\frac{C}{x}e^{-%
\frac{\sqrt{2\lambda }}{3}x^{3}}~.  \tag{4}
\end{equation}%
The constant $C$ in the asymptotic expression (4) is commonly referred to as
the asymptotic normalization constant of the wave function, and it clearly
depends on the parameter $\lambda $ --- $C=C\left( \lambda \right) $. The
asymptotic behavior of the function $\psi \left( x\right) $ as $x\rightarrow
-\infty $ can easily be expressed based on whether the wave function $\psi
\left( x\right) $ for the specified state is even or odd. Herein, due to the
parity of the Hamiltonian (1) with respect to the coordinate $x$, all
physical eigensolutions of equation (2) are divided into even and odd.

   The standard quantum mechanical approach to solving the 
 anharmonic oscillator problem involves applying perturbation theory
with respect to the parameter $\lambda $ [1, 4]. In this case, the
expansion of the standard Rayleigh-Schr\"{o}dinger perturbation theory for
the wave function $\psi \left( x\right) $ and for the ground state energy $%
E_{0}\left( \lambda \right) $ of this model can be expressed as typical
power series expansions around the point $\lambda =0$, in terms of the
coupling constant $\lambda $: 
\begin{equation}
\psi \left( \lambda ;x\right) =\overset{\infty }{\underset{n=0}{\dsum }}\chi
_{n}\left( x\right) \lambda ^{n}=\chi _{0}\left( x\right) +\chi _{1}\left(
x\right) \lambda +\chi _{2}\left( x\right) \lambda ^{2}+\ldots ~,  \tag{5}
\end{equation}%
\begin{equation}
E_{0}\left( \lambda \right) =\overset{\infty }{\underset{n=0}{\dsum }}%
A_{n}\lambda ^{n}=A_{0}+A_{1}\lambda +A_{2}\lambda ^{2}+\ldots ~,  \tag{6}
\end{equation}%
where the first coefficient $A_{0}=1/2$ of the latter expansion yields the
well-known value of the ground energy level of the standard harmonic
oscillator --- $E_{0}\left( 0\right) =\hbar \omega /2=1/2$. However, as
noted earlier, Bender and Wu [1] rigorously proved the
divergence of the perturbation theory series for the ground state energy $%
E_{0}\left( \lambda \right) $ of the anharmonic oscillator across the entire
complex plane of the coupling constant $\lambda $, with the sole exception
of the origin. In addition, for the anharmonic oscillator defined by
Hamiltonian (1), a number of qualitative arguments [21, 55, 56] have
been proposed that support the divergence of the perturbation theory series
for this model, among which the most important and well-known is the
so-called ``Dyson's instability argument'' [55]. The latter is also
sometimes referred to as ``Dyson's phenomenon'' or ``Dyson's effect,'' and
its essence lies in the qualitative change of the system's spectrum when the
sign of the parameter $\lambda $ is changed from positive to negative at any
arbitrarily small value of the constant $\lambda $. Specifically, when the
sign of $\lambda $ changes from positive to negative, the system's spectrum
``immediately'' becomes continuous and all discrete energy levels disappear.
Consequently, this demonstrates that the point $\lambda =0$ is a singular
point for energy levels as functions of $\lambda $, as a result of which the
Taylor power series, i.e., the perturbation theory series, diverge at this
point. A rigorous mathematical proof of this fact, first
qualitatively established by Dyson [55], 
and later thoroughly discussed, for example, in the review work [56], 
was provided by Bender and Wu in [1]. 
The mentioned divergence of the perturbation theory series for 
the anharmonic oscillator, as well as for many other perturbation series in
quantum mechanics and quantum field theory, not only holds significant
theoretical interest and importance, but also poses
substantial theoretical and practical difficulties for specific applications
and calculations. Consequently, practical implementation and 
calculations based on perturbation theory require the use of 
 specific methods and tools, such as special mathematical techniques
for summing divergent perturbation series. The latter presents a
considerable difficulty, especially with a relatively substantial increase
in the parameter $\lambda $, i.e., in the region of strong coupling.

   That said, there is yet another well-established approach 
to solving problems in quantum
mechanics [47--51], which
involves not expanding the system's wave function into  
the diverging perturbation series  
 (5) in terms of the coupling constant, but rather
expanding it in terms of the complete set of
eigenfunctions of the harmonic oscillator. In this case, the
zeroth-approximation Hamiltonian $H^{\left( 0\right) }$ corresponds to the
value of the coupling constant $\lambda =0$: 
\begin{equation}
H^{\left( 0\right) }=H\left( \lambda =0\right) =\frac{1}{2}\left(
p^{2}+x^{2}\right) ~,  \tag{7}
\end{equation}%
however, the expansion is not carried out in terms of the powers of the
parameter $\lambda $, but in terms of the eigenfunctions of the Hamiltonian
(7). At this point, $H^{\left( 0\right) }$ represents the Hamiltonian of the
 standard harmonic oscillator, and its eigenfunctions $\left\{
\varphi _{n}\left( x\right) \right\} _{n=0}^{\infty }$ are solutions to the
eigenvalue problem: 
\begin{equation}
H^{\left( 0\right) }\varphi _{n}\left( x\right) =E_{n}^{\left( 0\right)
}\varphi _{n}\left( x\right) ~,  \tag{8}
\end{equation}%
and the associated energy eigenvalues are given by
the well-known expression:%
\begin{equation}
E_{n}^{\left( 0\right) }=n+\frac{1}{2}~,~~~n=0,~1,~2,\ldots ~.  \tag{9}
\end{equation}%
Mutually orthogonal and normalized real eigenfunctions $\varphi _{n}\left(
x\right) $ of the harmonic oscillator are explicitly written as [57, 58]%
: 
\begin{equation}
\varphi _{n}\left( x\right) =\frac{1}{\sqrt{\sqrt{\pi }2^{n}n!}}e^{-\frac{%
x^{2}}{2}}H_{n}\left( x\right) ,~n=0,~1,~2,\ldots ,  \tag{10}
\end{equation}%
where $H_{n}\left( x\right) =\left( -1\right) ^{n}e^{x^{2}}\frac{d^{n}}{%
dx^{n}}e^{-x^{2}}$ are the classical Hermite polynomials, orthogonal with
the weight $w\left( x\right) =e^{-x^{2}}$ on the real axis $%
\mathbb{R}
=\left( -\infty ,+\infty \right) $. The eigenfunctions $\varphi _{n}\left(
x\right) $ here form a complete set, or basis, in the Hilbert space $%
L^{2}\left( 
\mathbb{R}
\right) $ of square-integrable functions, and the orthonormality condition
for them is expressed as 
\begin{equation}
\left\langle \varphi _{m}\mid \varphi _{n}\right\rangle =\left( \varphi
_{m},\varphi _{n}\right) =\tint\limits_{-\infty }^{\infty }\varphi
_{m}\left( x\right) \varphi _{n}\left( x\right) dx=\delta _{mn}~,  \tag{11}
\end{equation}%
where $\delta _{mn}$ are the usual Kronecker delta symbols ($\delta _{mn}=1$%
, $m=n$~; $\delta _{mn}=0$, $m\neq n$). It should be noted that due to the
respective properties of the Hermite polynomials, for even values of the
index $n$, the corresponding oscillator basis function $\varphi _{n}\left(
x\right) $ will be even in $x$, and for odd values of the index $n$, it will
accordingly be odd. The properties of the harmonic oscillator's wave
functions, as well as the properties of Hermite polynomials, are 
thoroughly studied, making many of them easy and convenient to use in
 work and applications. Thus, in particular, for the functions $%
\varphi _{n}\left( x\right) $, the following simple recurrence relations are
relevant [57]: 
\begin{equation}
x\varphi _{n}=\sqrt{\frac{n}{2}}\varphi _{n-1}+\sqrt{\frac{n+1}{2}}\varphi
_{n+1}~,  \tag{12}
\end{equation}%
which directly follow from the corresponding well-known recurrence relations
for Hermite polynomials. Multiplying the previous relation by $x$ and
applying it once more results in [57]{\ 
\begin{equation}
x^{2}\varphi _{n}=\frac{1}{2}\sqrt{n\left( n-1\right) }\varphi _{n-2}+\left(
n+\frac{1}{2}\right) \varphi _{n}+\frac{1}{2}\sqrt{\left( n+1\right) \left(
n+2\right) }\varphi _{n+2}~.  \tag{13}
\end{equation}%
The relationships (12), (13) will be employed in further work.}

   Since the wave function of our system, $\psi \left( x\right) $, is
square-integrable (see (3)), belongs to the $L^{2}\left( 
\mathbb{R}
\right) $ space, and also satisfies boundary conditions of decreasing
behavior at infinity similar to those of the oscillator basis functions, it
can therefore be expanded into a convergent functional series in terms of
the complete set of harmonic oscillator basis functions:{\ 
\begin{equation}
\psi \left( x\right) =\overset{\infty }{\underset{n=0}{\dsum }}c_{n}\varphi
_{n}\left( x\right) ~.  \tag{14}
\end{equation}%
Here, the real coefficients }$c_{n}${\ of this expansion, depending on} 
$\lambda${ as a parameter  } ($c_{n}=c_{n}\left( \lambda \right) $){,
satisfy the following normalization condition derived from (3) and (11):} 
\begin{equation}
\overset{\infty }{\underset{n=0}{\dsum }}c_{n}^{2}=1~.  \tag{15}
\end{equation}%
As will be seen later, the coefficients $c_{n}$ decrease relatively quickly
as $n$ increases, and the complete set of these coefficients $\left\{
c_{n}\right\} _{n=0}^{\infty }=\left\{ \left\langle \varphi _{n}\mid \psi
\right\rangle \right\} _{n=0}^{\infty }$ actually represents the wave
function of our system in the harmonic oscillator representation, or
oscillator representation, which is sometimes also referred to as the
``n''-representation. The last fact is illustrated by expressing these
coefficients in terms of the wave function $\psi $: 
\begin{equation}
c_{n}=\left\langle \varphi _{n}\mid \psi \right\rangle =\psi \left( n\right)
=\left( \varphi _{n},\psi \right) =\tint\limits_{-\infty }^{\infty }\varphi
_{n}\left( x\right) \psi \left( x\right) dx~.  \tag{16}
\end{equation}

   It is important to emphasize here that the rapid decrease of the
coefficients $c_{n}$ with increasing $n$, which in fact ensures the
convergence of the oscillator expansion (14), is fundamentally and
substantially different from the rapid factorial growth of the coefficients $%
A_{n}$ in the perturbation theory expansion (6) as $n$ increases. As a
consequence of this factorial growth of $A_{n}$ ($~A_{n}%
\underset{n\rightarrow \infty }{\sim }\Gamma \left( n+1/2\right) ~$),
established by Bender and Wu [1], the divergence of perturbation series
(6) becomes evident. Additionally, it should be noted that the
expansion in the oscillator basis (14) does not rely on the assumption of
the smallness of the perturbation $U$ and/or on power series expansions in
the parameter $\lambda $. In other words, for the expansion (14), the
qualitative arguments previously mentioned, which suggest the divergence of
the perturbation series in the parameter $\lambda $, do not apply and are
not valid. Indeed the functional series (14) converges for all 
 $x$, as will be seen later.
 This is also supported by the fact that the coefficients $%
c_{n}$ of this series decrease as $n$ increases, providing further 
evidence of this convergence.
 In general, the possibility of
expansion (14) for a square-integrable function $\psi \left( x\right) $ and
its convergence in the mean square sense in the $L^{2}\left( 
\mathbb{R}
\right) $ space, i.e., with respect to the norm of this space, follow from
the well-known completeness of the system of harmonic oscillator basis
functions, as demonstrated in quantum mechanics courses.

   Next, substituting our main expansion (14) into the initial Schr\"{o}%
dinger equation (2), we obtain in the usual manner an infinite homogeneous
system of linear algebraic equations for the coefficients $c_{n}$:{%
\ 
\begin{equation}
\overset{\infty }{\underset{n=0}{\dsum }}H_{mn}c_{n}=Ec_{m},~m=0,~1,~2,%
\ldots ~,  \tag{17}
\end{equation}%
where the matrix elements of the total Hamiltonian }$H${\ are provided by
expression: 
\begin{equation}
H_{mn}=\left\langle \varphi _{m}\mid H\mid \varphi _{n}\right\rangle =\left(
\varphi _{m},H\varphi _{n}\right) =\tint\limits_{-\infty }^{\infty }\varphi
_{m}\left( x\right) H\varphi _{n}\left( x\right) dx=H_{mn}^{\left( 0\right)
}+U_{mn}~.  \tag{18}
\end{equation}%
Here, the matrix elements of the zeroth-approximation Hamiltonian }$%
H^{\left( 0\right) }${  (7), which is the Hamiltonian of the harmonic
oscillator, are clearly diagonal in the eigenbasis and take the form:} 
\begin{equation}
H_{mn}^{\left( 0\right) }=\left( n+\frac{1}{2}\right) \delta _{mn}~. 
\tag{19}
\end{equation}%
Meanwhile, the matrix elements of the perturbation operator $U=\lambda V$
can be obtained using the recurrence relations (13), and for the elements $%
V_{mn}$, they can be written as [59, 60]  
\begin{equation*}
V_{mn}=\left\langle x^{4}\right\rangle _{mn}=\left\langle \varphi _{m}\mid
x^{4}\mid \varphi _{n}\right\rangle =\left( \varphi _{m},x^{4}\varphi
_{n}\right) =\left( x^{2}\varphi _{m},x^{2}\varphi _{n}\right) =
\end{equation*}%
\begin{equation*}
=\frac{\sqrt{n\left( n-1\right) \left( n-2\right) \left( n-3\right) }}{4}%
\delta _{m,n-4}+\frac{2n-1}{2}\sqrt{n\left( n-1\right) }\delta _{m,n-2}+%
\frac{3\left( 2n^{2}+2n+1\right) }{4}\delta _{mn} 
\end{equation*}%
\begin{equation}
+\frac{2n+3}{2}\sqrt{\left( n+1\right) \left( n+2\right) }\delta _{m,n+2}+%
\frac{\sqrt{\left( n+1\right) \left( n+2\right) \left( n+3\right) \left(
n+4\right) }}{4}\delta _{m,n+4}~.  \tag{20}
\end{equation}%
In this case, the relation $U_{mn}=\lambda V_{mn}$ is also clearly valid.

   For a specific numerical solution of the obtained infinite system of linear
algebraic equations (17), with its matrix elements explicitly given by
(18)--(20), it is necessary to truncate the expansion (14) and,
consequently, the system (17) to a finite number of terms $M$. Thus,
substituting the obtained matrix elements into Eq. (17), we can then
write the system of homogeneous linear algebraic equations for determining
the coefficients $c_{n}$ in the following finite form: 
\begin{equation}
\overset{M}{\underset{n=0}{\dsum }}\left[ \left( n+\frac{1}{2}\right) \delta
_{mn}+\lambda V_{mn}\right] c_{n}=Ec_{m},~m=0,~1,~2,\ldots ,M~,  \tag{21}
\end{equation}%
where the matrix elements $V_{mn}$ are given explicitly by formula (20). The
system of equations (21) can also be rewritten in the formal matrix form, as
is customary in matrix quantum mechanics: 
\begin{equation}
\mathbf{H}\Psi =E\Psi ~.  \tag{22}
\end{equation}%
Here, $\mathbf{H=}\left[ H_{mn}\right] $ is a finite real symmetric
Hamiltonian matrix of dimension $\left( M+1\right) \times \left( M+1\right) $
with the elements 
\begin{equation}
H_{mn}=\left( n+\frac{1}{2}\right) \delta _{mn}+\lambda
V_{mn},~m,n=0,~1,~2,\ldots ,M~,  \tag{23}
\end{equation}%
and $\Psi $ is a finite column vector of coefficients $c_{n}$: 
\begin{equation}
\Psi =\left[ 
\begin{array}{c}
c_{0} \\ 
c_{1} \\ 
\vdots \\ 
c_{M}%
\end{array}%
\right] ~,  \tag{24}
\end{equation}%
which actually represents the wave function in matrix representation. In
this context, the system's energy eigenvalues $E_{n}$ are the eigenvalues of
the matrix $\mathbf{H}$, and the corresponding eigenvectors are $\Psi _{n}$,
which, given the normalization condition (15), must be chosen as unit
vectors, i.e., the following condition must be satisfied: 
\begin{equation}
\left\vert \left\vert \Psi \right\vert \right\vert
^{2}=c_{0}^{2}+c_{1}^{2}+\ldots +c_{M}^{2}=1~.  \tag{25}
\end{equation}

   For the final numerical calculations of the properties of specific physical
states of the anharmonic oscillator, one must also 
account for the parity of these states. Thus, for the ground state of the system
and other even states, the expansion (14) in terms of the oscillator basis
functions will contain only even basis functions, with $c_{2j+1}=0$ for even
states. By setting $m=2i$ and $n=2j$ in (21), and taking into account the
even order of the system of equations with $M=2N$, we can finally write the
system of homogeneous linear algebraic equations for determining the
coefficients $c_{2j}$ in the case of even states of the system as follows: 
\begin{equation}
\overset{N}{\underset{j=0}{\dsum }}\left[ \left( 2j+\frac{1}{2}\right)
\delta _{ij}+\lambda V_{2i,2j}\right] c_{2j}=Ec_{2i},~i=0,~1,~2,\ldots ,N~. 
\tag{26}
\end{equation}%
A similar system of equations for the case of odd states of the anharmonic
oscillator ($M=2N+1$ ) is given by: 
\begin{equation}
\overset{N}{\underset{j=0}{\dsum }}\left[ \left( 2j+\frac{3}{2}\right)
\delta _{ij}+\lambda V_{2i+1,2j+1}\right] c_{2j+1}=Ec_{2i+1},~i=0,~1,~2,%
\ldots ,N~.  \tag{27}
\end{equation}%
In the latter case, the expansion (14) in terms of the oscillator basis
functions will include only odd basis functions, meaning that $c_{2j}=0$ for
odd states. 

    Both systems (26) and (27) can be written in the same matrix form (22), 
where the corresponding real symmetric Hamiltonian matrix has 
dimension $\left( N+1\right) \times \left( N+1\right) $, and its 
eigenvalues and eigenvectors are to be found.
It should be particularly
emphasized that determining the eigenvalues and eigenvectors
of a finite-order real symmetric square matrix, as described in (22), is
a classical problem, with highly developed solution techniques in
computational linear algebra. 
Numerous efficient numerical algorithms and readily available subroutines 
can be effectively used to carry out this computation in practical applications. 
This task is, in essence, equivalent to diagonalizing the given matrix. 
Thus, the problem of finding the eigenenergy levels and their corresponding
eigenwave functions for the quartic anharmonic oscillator is reduced in this
approach to determining the eigenvalues and eigenvectors of a
finite real symmetric square matrix of the system's Hamiltonian in the
harmonic oscillator representation (the ``n''-representation), with matrix
elements given by formulas (23) and (20), also taking into account the
parity of the corresponding states. The system's wave function, $\psi \left(
x\right) $, in the coordinate representation, considering also its parity,
will be given by the expansion (14) over the complete set of harmonic
oscillator basis functions, retaining a finite number of $N+1$
terms ($n=\overline{0,N}$ ). 
Subsequent calculations and convergence tests will demonstrate that 
the computed eigenvalues (i.e., energy levels of the anharmonic oscillator) 
converge well with increasing $N$, as obtained by solving the eigenvalue 
problem for the truncated system (21), derived from (17). 
\vspace*{0.5em}

\begin{center}
{\textbf{ 3. Ground State Energy: Calculation Results and Convergence Analysis }}  
\end{center}
In Table 1, for various values of $\lambda $, the results of calculations
for the eigenvalues of the ground state energy $E_{0}\left( \lambda \right) $
of the quartic anharmonic oscillator are presented with a standard precision
of eight significant digits after the decimal point. These values are calculated
at various orders of the expansion (14) of the system's wave function over
the oscillator basis, as per the method discussed above, i.e., specifically
as eigenvalues of the system of equations (26). Herein, the values of the
coupling constant $\lambda $ were chosen in the weak coupling region $%
\lambda \lesssim 1$. The results demonstrate, in this case,  
an excellent convergence rate and very high achievable accuracy of
calculations when employing oscillator expansions of very low order $%
N\lesssim 15$ in this region. Also, with the increase in the coupling
constant $\lambda $, a slowdown in the convergence rate of the calculated
values becomes quite evident, 
which is entirely expected, since   
as $\lambda $ increases, the
 \textquotedblleft strength\textquotedblright\ of the 
 perturbation defined by operator $U\left(
x\right) =\lambda x^{4}$ increases, and consequently, the \textquotedblleft
deviation\textquotedblright\ of the system from a pure harmonic oscillator
grows. As a result, to achieve a good description and convergence, it
becomes necessary to include an increasing number of oscillator basis
functions (harmonic oscillator \textquotedblleft quanta\textquotedblright )
in the expansion.
\begin{table}[h] 
\captionsetup{width=0.91\textwidth} 
{\boldmath{ 
\caption{\footnotesize \textbf{ The ground state energy values $E_{0}$
of the quartic anharmonic oscillator, calculated in various orders $N$ of
oscillator basis expansion for certain values of the coupling constant $%
\lambda $ in the weak coupling region $\lambda \lesssim 1$ (number of basis
functions used in the expansion is $N+1$). The last row labeled
\textquotedblleft Ex.\textquotedblright\ (\textquotedblleft
Exact\textquotedblright ) presents the exact values.}} 
}} 
\vspace*{-1.5em}  
 \begin{center} 
{  
\setlength{\tabcolsep}{5pt} 
{\footnotesize  
\begin{tabular}{@{}cccccccc@{}} 
\hline\hline   
$N$ & $\lambda =0.01$ & $\lambda =0.05$ & $\lambda =0.1$ & $\lambda =0.25$ & 
$\lambda =0.3$ & $\lambda =0.5$ & $\lambda =0.7$ \\ \hline  
1 & 0.50728471 & 0.53291674 & 0.55956491 & 0.62232307 & 0.64035424 & 
0.70630142 & 0.76733622 \\ 
2 & 0.50725644 & 0.53268783 & 0.55938559 & 0.62186431 & 0.63906797 & 
0.69753577 & 0.74582288 \\ 
3 & 0.50725621 & 0.53264342 & 0.55916595 & 0.62129174 & 0.63853982 & 
0.69745351 & 0.74566724 \\ 
4 & 0.50725620 & 0.53264285 & 0.55914665 & 0.62097863 & 0.63809736 & 
0.69668553 & 0.74496964 \\ 
5 & 0.50725620 & 0.53264278 & 0.55914661 & 0.62092968 & 0.63800011 & 
0.69628302 & 0.74425930 \\ 
6 & 0.50725620 & 0.53264276 & 0.55914640 & 0.62092802 & 0.63799291 & 
0.69618747 & 0.74397538 \\ 
7 & 0.50725620 & 0.53264275 & 0.55914633 & 0.62092772 & 0.63799287 & 
0.69617775 & 0.74391212 \\ 
8 & 0.50725620 & 0.53264275 & 0.55914633 & 0.62092723 & 0.63799225 & 
0.69617774 & 0.74390602 \\ 
9 & 0.50725620 & 0.53264275 & 0.55914633 & 0.62092706 & 0.63799188 & 
0.69617696 & 0.74390601 \\ 
10 & 0.50725620 & 0.53264275 & 0.55914633 & 0.62092703 & 0.63799179 & 
0.69617622 & 0.74390508 \\ 
11 & 0.50725620 & 0.53264275 & 0.55914633 & 0.62092703 & 0.63799178 & 
0.69617591 & 0.74390415 \\ 
12 & 0.50725620 & 0.53264275 & 0.55914633 & 0.62092703 & 0.63799178 & 
0.69617583 & 0.74390368 \\ 
13 & 0.50725620 & 0.53264275 & 0.55914633 & 0.62092703 & 0.63799178 & 
0.69617582 & 0.74390353 \\ 
14 & 0.50725620 & 0.53264275 & 0.55914633 & 0.62092703 & 0.63799178 & 
0.69617582 & 0.74390350 \\ 
15 & 0.50725620 & 0.53264275 & 0.55914633 & 0.62092703 & 0.63799178 & 
0.69617582 & 0.74390350 \\ \hline  
Ex. & 0.50725620 & 0.53264275 & 0.55914633 & 0.62092703 & 0.63799178 & 
0.69617582 & 0.74390350 \\ \hline\hline  
\end{tabular} }  
\label{tab1} }
 \end{center} 
\end{table}  
  
   In Table 2, the calculated energy values $E_{0}\left( \lambda \right) $ of
the ground state of the anharmonic oscillator for certain values of the
coupling constant $\lambda $ from the intermediate and strong coupling
region $1\lesssim \lambda \lesssim 50$ are presented. The results in Table 2
show a further, predictable deterioration in the convergence of the calculation
results as the values of the coupling constant $\lambda $ increase.
Nevertheless, the proposed calculation method enables reliable and precise
computations across a relatively wide range of values of the coupling constant $%
\lambda \lesssim 50$, employing oscillator expansions and the subsequent
diagonalization of square matrices of relatively low order $N\lesssim 100$. 
\begin{table}[h] 
\captionsetup{width=0.89\textwidth} 
{\boldmath{ 
\caption{\footnotesize \textbf{The ground state energy values $E_{0}$ 
of the quartic anharmonic oscillator, calculated in various orders $N$ of
oscillator basis expansion for certain values of the coupling constant 
$\lambda $ in the intermediate and strong coupling region $1\lesssim \lambda
\lesssim 50$ (number of basis functions used in the expansion is $N+1$).}} 
}}   
\vspace*{-1.5em}  
 \begin{center} 
{
\setlength{\tabcolsep}{5pt} 
{\footnotesize 
\begin{tabular}{@{}cccccccc@{}} 
\hline\hline
$N$ & $\lambda =1$ & $\lambda =2$ & $\lambda =5$ & $\lambda =10$ & $\lambda
=20$ & $\lambda =25$ & $\lambda =50$ \\ \hline
5 & 0.80469834 & 0.95435205 & 1.22955456 & 1.53280435 & 2.02335509 & 
2.25295532 & 3.36402214 \\ 
10 & 0.80377399 & 0.95159076 & 1.22582543 & 1.50976951 & 1.87313929 & 
2.01046262 & 2.53806547 \\ 
15 & 0.80377068 & 0.95157068 & 1.22459892 & 1.50543697 & 1.86945103 & 
2.00761035 & 2.51003367 \\ 
20 & 0.80377065 & 0.95156849 & 1.22459162 & 1.50498451 & 1.86608440 & 
2.00297807 & 2.50597337 \\ 
25 & 0.80377065 & 0.95156848 & 1.22458721 & 1.50497950 & 1.86571365 & 
2.00204840 & 2.50125468 \\ 
30 & 0.80377065 & 0.95156847 & 1.22458705 & 1.50497317 & 1.86570865 & 
2.00201274 & 2.49988406 \\ 
35 & 0.80377065 & 0.95156847 & 1.22458704 & 1.50497243 & 1.86569949 & 
2.00200522 & 2.49973169 \\ 
40 & 0.80377065 & 0.95156847 & 1.22458704 & 1.50497242 & 1.86569616 & 
2.00199816 & 2.49972861 \\ 
45 & 0.80377065 & 0.95156847 & 1.22458704 & 1.50497241 & 1.86569583 & 
2.00199653 & 2.49971903 \\ 
50 & 0.80377065 & 0.95156847 & 1.22458704 & 1.50497241 & 1.86569583 & 
2.00199642 & 2.49971167 \\ 
55 & 0.80377065 & 0.95156847 & 1.22458704 & 1.50497241 & 1.86569581 & 
2.00199641 & 2.49970923 \\ 
60 & 0.80377065 & 0.95156847 & 1.22458704 & 1.50497241 & 1.86569580 & 
2.00199639 & 2.49970883 \\ 
65 & 0.80377065 & 0.95156847 & 1.22458704 & 1.50497241 & 1.86569580 & 
2.00199639 & 2.49970881 \\ 
70 & 0.80377065 & 0.95156847 & 1.22458704 & 1.50497241 & 1.86569580 & 
2.00199638 & 2.49970880 \\ 
75 & 0.80377065 & 0.95156847 & 1.22458704 & 1.50497241 & 1.86569580 & 
2.00199638 & 2.49970879 \\ 
80 & 0.80377065 & 0.95156847 & 1.22458704 & 1.50497241 & 1.86569580 & 
2.00199638 & 2.49970878 \\ 
85 & 0.80377065 & 0.95156847 & 1.22458704 & 1.50497241 & 1.86569580 & 
2.00199638 & 2.49970877 \\ \hline
Ex. & 0.80377065 & 0.95156847 & 1.22458704 & 1.50497241 & 1.86569580 & 
2.00199638 & 2.49970877 \\ \hline\hline
\end{tabular} }  
\label{tab2} }
 \end{center} 
\end{table}

   In Table 3, the energy values $E_{0}\left( \lambda \right) $ of the ground
state of the anharmonic oscillator are presented for certain values of the
coupling constant $\lambda $ from the superstrong coupling region $\lambda
\gtrsim 100$, also calculated using the proposed method. It is evident that
in this region, as the values of the coupling constant $\lambda $ increase,
the convergence of the calculation results deteriorates significantly, and
to achieve the required accuracy, it is necessary to take into account
several hundred oscillator basis functions in the expansion. However, in
general, choosing an expansion order $N\lesssim 700$ is sufficient. In this
case, it is also important to highlight that the aforementioned
well-developed methods for solving the problem of finding eigenvalues and
eigenvectors of square matrices make it entirely feasible to work with and
manipulate matrices of very large dimensions, up to $N\sim 10000\div
100000$, if necessary for solving this problem. But overall, for
calculations with a standard precision of eight significant digits after the
decimal point in the practically necessary range of the coupling constant $%
\lambda $, choosing an expansion order $N\lesssim 1000$ is entirely
sufficient. 
\begin{table}[h] 
{\boldmath{ 
\caption{\footnotesize \textbf{The ground state energy values $E_{0}$
of the quartic anharmonic oscillator, calculated in various orders $N$ of
oscillator basis expansion for certain values of the coupling constant 
$\lambda $ in the superstrong coupling region $\lambda \gtrsim 100$
(number of basis functions used in the expansion is $N+1$).}}  
}} 
\vspace*{-1.5em} 
 \begin{center} 
{ 
\setlength{\tabcolsep}{5pt} 
{\footnotesize 
\begin{tabular}{@{}cccccccc@{}}
\hline\hline
$N$ & $\lambda =100$ & $\lambda =500$ & $\lambda =1000$ & $\lambda =2000$ & $%
\lambda =5000$ & $\lambda =10000$ & $\lambda =20000$ \\ \hline
50 & 3.13141038 & 5.32768645 & 6.71975732 & 8.46418574 & 11.68508581 & 
15.80752057 & 23.37875096 \\ 
100 & 3.13138417 & 5.31991445 & 6.69428156 & 8.42901540 & 11.45877776 & 
14.45908609 & 18.23396405 \\ 
150 & 3.13138416 & 5.31989443 & 6.69422256 & 8.42755115 & 11.43128071 & 
14.40965840 & 18.19404951 \\ 
200 & 3.13138416 & 5.31989436 & 6.69422091 & 8.42749872 & 11.43088535 & 
14.39824368 & 18.14601842 \\ 
250 & 3.13138416 & 5.31989436 & 6.69422085 & 8.42749826 & 11.43081004 & 
14.39810218 & 18.13759091 \\ 
300 & 3.13138416 & 5.31989436 & 6.69422085 & 8.42749818 & 11.43080463 & 
14.39801275 & 18.13738649 \\ 
350 & 3.13138416 & 5.31989436 & 6.69422085 & 8.42749818 & 11.43080451 & 
14.39799591 & 18.13729054 \\ 
400 & 3.13138416 & 5.31989436 & 6.69422085 & 8.42749818 & 11.43080444 & 
14.39799558 & 18.13723704 \\ 
450 & 3.13138416 & 5.31989436 & 6.69422085 & 8.42749818 & 11.43080444 & 
14.39799541 & 18.13722954 \\ 
500 & 3.13138416 & 5.31989436 & 6.69422085 & 8.42749818 & 11.43080444 & 
14.39799535 & 18.13722937 \\ 
550 & 3.13138416 & 5.31989436 & 6.69422085 & 8.42749818 & 11.43080444 & 
14.39799534 & 18.13722920 \\ 
600 & 3.13138416 & 5.31989436 & 6.69422085 & 8.42749818 & 11.43080444 & 
14.39799534 & 18.13722909 \\ 
650 & 3.13138416 & 5.31989436 & 6.69422085 & 8.42749818 & 11.43080444 & 
14.39799534 & 18.13722907 \\ 
700 & 3.13138416 & 5.31989436 & 6.69422085 & 8.42749818 & 11.43080444 & 
14.39799534 & 18.13722907 \\ \hline
Ex. & 3.13138416 & 5.31989436 & 6.69422085 & 8.42749818 & 11.43080444 & 
14.39799534 & 18.13722907 \\ \hline\hline
\end{tabular} } 
\label{tab3} }
 \end{center} 
\end{table}

 \newpage
   However, more recently, there has been a growing interest and practical
importance in what are known as high-precision calculations of the physical
characteristics for different models of the anharmonic 
oscillator [9, 36, 41, 42], that is, calculations performed 
with an accuracy of several tens of 
significant digits. For example, in a recent work [41], quantities
characterizing the oscillator, or more precisely, a double-well
potential related to it, were calculated with an accuracy of forty
significant digits. The justification for this interest and the significance
of high-precision calculations can be found in the aforementioned studies.
For the execution of such high-precision calculations within the framework
of the approach being discussed, the use of oscillator expansions and the
corresponding diagonalization of matrices of increasingly higher order are
clearly necessary. Given the previously mentioned development of methods for
solving this problem, all of this is entirely feasible. In Table 4, the
results of high-precision calculations with an accuracy of twenty
significant digits after the decimal point are presented for the ground state
energy values $E_{0}\left( \lambda \right) $ of the oscillator in the case
of certain values of the parameter $\lambda $ from the region of
weak and intermediate coupling $\lambda \lesssim 50$. From Table 4, it is
apparent that achieving this level of accuracy requires increasing the
dimension of the oscillator basis utilized by approximately a factor of three 
compared to the basis previously used. And overall, it is evident that
high-precision calculations are entirely achievable within the framework of
the oscillator basis expansion approach under consideration. 
\begin{table}[h] 
\captionsetup{width=0.85\textwidth} 
{\boldmath{ 
\caption{\footnotesize \textbf{High-precision calculation results for
the ground state energy $E_{0}$ values of the quartic anharmonic
oscillator with an accuracy of 20 significant digits after the decimal,
given for certain values of the coupling constant $\lambda $. $N$ is
the minimal order of expansion in the oscillator basis required to achieve
this degree of accuracy.}}  
}} 
\vspace*{-1.5em} 
 \begin{center} 
{ 
\setlength{\tabcolsep}{22pt} 
{\footnotesize 
\begin{tabular}{@{}ccc@{}}
\hline\hline
$\lambda $ & $N$ & $E_{0}$ \\ \hline
0.1 & 25 & 0.55914632718351957672 \\ 
0.25 & 34 & 0.62092702982574866086 \\ 
0.5 & 46 & 0.69617582076514592783 \\ 
1 & 61 & 0.80377065123427376935 \\ 
5 & 106 & 1.22458703605919345913 \\ 
10 & 134 & 1.50497240777889109916 \\ 
25 & 177 & 2.00199638413881390772 \\ 
50 & 224 & 2.49970877256879391465 \\ \hline\hline
\end{tabular} } 
\label{tab4} }
 \end{center} 
\end{table} 

\begin{center}
{\textbf{ 4. Excited States: Calculation Results and Convergence Analysis }}  
\end{center}
{\noindent The study and calculation of the energies of excited states of the
anharmonic oscillator are of great interest both from a theoretical
standpoint and in terms of practical applications and implications for
physical models of various quantum systems. In this context, the
calculations of energy levels and their corresponding wave functions within
the studied approach should be performed for even states based on the
solution of the system of equations (26), and for odd states based on the
solution of the system of equations (27). Practical calculations show that
the order of the oscillator basis required for calculating the energies of
excited states should be somewhat increased compared to the corresponding
order for the ground state, i.e., at the same value of the coupling constant 
$\lambda $. However, this necessary increase in the expansion 
order is not drastic but 
quite moderate, as evidenced by the calculated energy values of the first
excited state $E_{1}\left( \lambda \right) $ of the quartic anharmonic
oscillator for the weak coupling region, presented in Table 5. In this case,
to achieve standard precision, it is sufficient to take into account about
20 terms ($N\lesssim 20$) in the oscillator expansions, compared to 15 terms
($N\lesssim 15$) for calculating the ground state energy in this range of 
$\lambda $ values. Thus, a relatively small
increase in the order of the employed oscillator basis enables reliable
calculations of the energy levels of a whole series of the lowest excited
states of the anharmonic oscillator.}
\begin{table}[h] 
\captionsetup{width=0.97\textwidth}
{\boldmath{ 
\caption{\footnotesize \textbf{The first excited state energy values 
$E_{1}$ of the quartic anharmonic oscillator, calculated in various orders $N$
of oscillator basis expansion for certain values of the coupling constant 
$\lambda $ in the weak coupling region $\lambda \lesssim 1$ (number of basis
functions used in the expansion is $N+1$).}}  
}} 
\vspace*{-1.5em} 
 \begin{center} 
{ 
\setlength{\tabcolsep}{5pt} 
{\footnotesize 
\begin{tabular}{@{}cccccccc@{}}
\hline\hline
$N$ & $\lambda =0.01$ & $\lambda =0.05$ & $\lambda =0.1$ & $\lambda =0.25$ & 
$\lambda =0.3$ & $\lambda =0.5$ & $\lambda =0.7$ \\ \hline
1 & 1.53575723 & 1.65382154 & 1.77095038 & 2.05529644 & 2.14170387 & 
2.47367272 & 2.79624680 \\ 
2 & 1.53565077 & 1.65370349 & 1.77032414 & 2.02739820 & 2.09685759 & 
2.33813960 & 2.54972832 \\ 
3 & 1.53564829 & 1.65345168 & 1.76971379 & 2.02736986 & 2.09628152 & 
2.32643730 & 2.51326463 \\ 
4 & 1.53564828 & 1.65343619 & 1.76951559 & 2.02652229 & 2.09552229 & 
2.32643450 & 2.51171174 \\ 
5 & 1.53564828 & 1.65343618 & 1.76950320 & 2.02606063 & 2.09485084 & 
2.32546932 & 2.51127708 \\ 
6 & 1.53564828 & 1.65343603 & 1.76950316 & 2.02597275 & 2.09466523 & 
2.32471990 & 2.51019723 \\ 
7 & 1.53564828 & 1.65343601 & 1.76950276 & 2.02596784 & 2.09464413 & 
2.32446019 & 2.50952338 \\ 
8 & 1.53564828 & 1.65343601 & 1.76950265 & 2.02596760 & 2.09464402 & 
2.32441214 & 2.50928581 \\ 
9 & 1.53564828 & 1.65343601 & 1.76950264 & 2.02596670 & 2.09464315 & 
2.32440937 & 2.50923596 \\ 
10 & 1.53564828 & 1.65343601 & 1.76950264 & 2.02596627 & 2.09464234 & 
2.32440903 & 2.50923189 \\ 
11 & 1.53564828 & 1.65343601 & 1.76950264 & 2.02596617 & 2.09464204 & 
2.32440774 & 2.50923174 \\ 
12 & 1.53564828 & 1.65343601 & 1.76950264 & 2.02596617 & 2.09464199 & 
2.32440682 & 2.50923039 \\ 
13 & 1.53564828 & 1.65343601 & 1.76950264 & 2.02596617 & 2.09464199 & 
2.32440647 & 2.50922909 \\ 
14 & 1.53564828 & 1.65343601 & 1.76950264 & 2.02596617 & 2.09464199 & 
2.32440637 & 2.50922841 \\ 
15 & 1.53564828 & 1.65343601 & 1.76950264 & 2.02596616 & 2.09464199 & 
2.32440636 & 2.50922817 \\ 
16 & 1.53564828 & 1.65343601 & 1.76950264 & 2.02596616 & 2.09464199 & 
2.32440636 & 2.50922811 \\ 
17 & 1.53564828 & 1.65343601 & 1.76950264 & 2.02596616 & 2.09464199 & 
2.32440636 & 2.50922811 \\ 
18 & 1.53564828 & 1.65343601 & 1.76950264 & 2.02596616 & 2.09464199 & 
2.32440635 & 2.50922811 \\ 
19 & 1.53564828 & 1.65343601 & 1.76950264 & 2.02596616 & 2.09464199 & 
2.32440635 & 2.50922811 \\ 
20 & 1.53564828 & 1.65343601 & 1.76950264 & 2.02596616 & 2.09464199 & 
2.32440635 & 2.50922810 \\ \hline
Ex. & 1.53564828 & 1.65343601 & 1.76950264 & 2.02596616 & 2.09464199 & 
2.32440635 & 2.50922810 \\ \hline\hline
\end{tabular} }  
\label{tab5} }
 \end{center} 
\end{table} 

   In Table 6, the calculated energy values 
of the first six excited levels $E_{1}$--$E_{6}$ of the quartic anharmonic 
oscillator are presented for a wide range of values of the 
coupling constant $\lambda $,  
 extensively covering the entire interval for possible practical applications. 
Therefore, Table 6 demonstrates the capability of the considered method to reliably
obtain energy values of a wide spectrum of excited oscillator states across
a broad range of variation of the parameter $\lambda $, utilizing an oscillator
function basis of moderately low dimension $N\lesssim 1000$.  
\begin{table}[h] 
\captionsetup{width=0.90\textwidth} 
{\boldmath{ 
\caption{\footnotesize \textbf{The energy values $E_{1}$--$E_{6}$ for
the first six excited states of the quartic anharmonic oscillator at certain
values of the coupling constant $\lambda $, calculated according to the
described method of expanding the system's wave function into the oscillator
basis.}}  
}} 
\vspace*{-1.5em} 
 \begin{center} 
{
\setlength{\tabcolsep}{5pt} 
{\footnotesize 
\begin{tabular}{@{}lcccccc@{}}
\hline\hline
$\lambda $ & {$E_{1}$} & {$E_{2}$} & {$E_{3}$} & {$E_{4}$} & {$E_{5}$} & {$%
E_{6}$} \\ \hline
0.01 & 1.53564828 & 2.59084580 & 3.67109494 & 4.77491312 & 5.90102667 & 
7.04832688 \\ 
0.05 & 1.65343601 & 2.87397963 & 4.17633891 & 5.54929781 & 6.98496310 & 
8.47739734 \\ 
0.1 & 1.76950264 & 3.13862431 & 4.62888281 & 6.22030090 & 7.89976723 & 
9.65783999 \\ 
0.25 & 2.02596616 & 3.69845032 & 5.55757714 & 7.56842287 & 9.70914788 & 
11.96454362 \\ 
0.3 & 2.09464199 & 3.84478265 & 5.79657363 & 7.91175273 & 10.16648889 & 
12.54425866 \\ 
0.5 & 2.32440635 & 4.32752498 & 6.57840195 & 9.02877872 & 11.64872073 & 
14.41766923 \\ 
0.7 & 2.50922810 & 4.71032810 & 7.19326528 & 9.90261070 & 12.80392971 & 
15.87368362 \\ 
1 & 2.73789227 & 5.17929169 & 7.94240398 & 10.96358309 & 14.20313910 & 
17.63404912 \\ 
2 & 3.29286782 & 6.30388057 & 9.72732317 & 13.48127584 & 17.51413240 & 
21.79095639 \\ 
5 & 4.29950173 & 8.31796075 & 12.90313811 & 17.94258561 & 23.36454045 & 
29.12064937 \\ 
10 & 5.32160826 & 10.34705559 & 16.09014687 & 22.40875129 & 29.21148486 & 
36.43690897 \\ 
20 & 6.62845235 & 12.93046099 & 20.13941486 & 28.07599084 & 36.62427661 & 
45.70645455 \\ 
25 & 7.12085350 & 13.90198143 & 21.66077524 & 30.20401603 & 39.40664348 & 
49.18472735 \\ 
50 & 8.91509636 & 17.43699213 & 27.19264579 & 37.93850201 & 49.51641866 & 
61.82034881 \\ 
100 & 11.18725425 & 21.90689815 & 34.18252411 & 47.70720589 & 62.28123797 & 
77.77077060 \\ 
500 & 19.04341673 & 37.34070210 & 58.30159947 & 81.40118710 & 106.29709170 & 
132.75997584 \\ 
1000 & 23.97220606 & 47.01733873 & 73.41911384 & 102.51615713 & 133.87689122
& 167.21225819 \\ 
2000 & 30.18641645 & 59.21511396 & 92.47347155 & 129.12819984 & 168.63537539
& 210.63072012 \\ 
5000 & 40.95165848 & 80.34295631 & 125.47537195 & 175.21794811 & 228.83228750
& 285.82389581 \\ 
10000 & 51.58610333 & 101.21231580 & 158.07220754 & 220.74085668 & 
288.28784144 & 360.09008661 \\ 
20000 & 64.98667570 & 127.50883864 & 199.14512348 & 278.10023732 & 
363.20184322 & 453.66487479 \\ \hline\hline
\end{tabular} } 
\label{tab6} }
 \end{center} 
\end{table}

\begin{center}
{\textbf{ 5. Wave Functions: Computation and Convergence Analysis }}  
\end{center}
{\noindent Thus, to study the model of the quartic anharmonic oscillator, 
we proposed 
using the method of expanding the system's wave function in terms of the
complete set of the harmonic oscillator eigenfunctions. Utilizing this
approach, the energy values of the ground and the first six excited levels
of the anharmonic oscillator were calculated across a broad range of 
variation of the
oscillator's coupling constant $\lambda $, through a
diagonalization process of the system's Hamiltonian in its oscillator
\textquotedblleft n\textquotedblright -representation. Next, we will  
consider the important issue of constructing and calculating the wave
functions of the ground and excited states of the quartic anharmonic
oscillator using the proposed method. In the studied approach, the wave
function of the system $\psi \left( x\right) $ in the coordinate
representation is given by the main convergent expansion (14) over the oscillator
basis, which includes a certain finite number of terms in accordance with
the chosen expansion order $N$.  As stated above, it is necessary to
take into account the parity of the state under consideration. Thus, for the
ground and other even states of the system, only even harmonic oscillator
basis functions appear in expansion (14), i.e., only the even coefficients $%
c_{2n}$ of this expansion are nonzero --- $\left\{ c_{2n}\right\} _{n=0}^{N}$%
. These coefficients $c_{2n}$ are determined, at a given $\lambda $, by
solving a system of homogeneous linear algebraic equations (26), i.e., as
coordinates of unit-normalized eigenvectors of the finite, real, symmetric
Hamiltonian matrix of the system with elements (23), corresponding to a
specific energy eigenvalue already found.} 

   In Table 7, the values of
the coefficients $c_{2n}$ for the oscillator's ground state, in the case of
two typical values of the coupling constant --- $\lambda =1$ and $\lambda
=10 $, are presented, calculated as described above. In doing so, the order
of expansion in the oscillator basis was chosen to be $N=89$, which
corresponds to the inclusion of 90 basis functions in the main expansion
(14). Recall that these coefficients $c_{2n}$ are essentially the wave
function of the system in the oscillator representation. The results in
Table 7 demonstrate notably good and rapid convergence of the calculated
coefficients depending on their order. In this case, the decrease in the
convergence rate of the coefficients $c_{2n}$ as 
$\lambda $ increases is relatively moderate. Using 90 basis functions for the
expansion, the accuracy of the calculated coefficients $c_{2n}$, i.e., the
wave function in the oscillator representation, reaches $\sim
10^{-16}$ for $\lambda =1$ and $\sim 10^{-10}$ for $\lambda =10$. In fact,
Table 7 presents the tabulated wave function of the system for these two 
values of the coupling constant, and any specific value of the wave function $\psi
_{0}\left( \lambda ;x\right) $ in the coordinate representation, at a given
value of the coordinate $x$, can be easily obtained from the main expansion
(14). However, it should be noted that Table 7 provides only eight
significant digits after the decimal point for the calculated values of the
coefficients $c_{2n}$. A more accurate calculation is achieved by taking
into account a correspondingly larger number of significant digits when
computing the coordinates of the unit-normalized eigenvectors 
 of the Hamiltonian (23). 
\begin{table}[H] 
\captionsetup{width=0.96\textwidth}
{\boldmath{ 
\caption{\footnotesize \textbf{ {
Values of the coefficients $c_{2n}$ 
for the expansion (14) in the oscillator basis of the ground state wave
function $\psi _{0}\left( \lambda ;x\right) $ of the quartic anharmonic
oscillator for the coupling constants  $\lambda =1$ and 
$\lambda=10 $. The expansion order in the oscillator basis is chosen as 
$N=89 $ (accounting for 90 basis functions).}} }  
}} 
\vspace*{-1.2em} 
 \begin{center} 
{ 
\setlength{\tabcolsep}{2pt}
{\footnotesize   
\begin{tabular}{@{}ccc|ccc|ccc@{}}
\hline\hline
$n$ & $c_{2n}\left( \lambda =1\right) $ & $c_{2n}\left( \lambda =10\right) $
& $n$ & $c_{2n}\left( \lambda =1\right) $ & $c_{2n}\left( \lambda =10\right) 
$ & $n$ & $c_{2n}\left( \lambda =1\right) $ & $c_{2n}\left( \lambda
=10\right) $ \\ \hline
0 & 9.70795971e-1 & 8.92672723e-1 & 30 & 6.25141823e-8 & 3.26660738e-5 & 60
& -1.86029470e-12 & 1.16871292e-7 \\ 
1 & -2.33973361e-1 & -3.87986676e-1 & 31 & -3.36708449e-8 & -2.10863726e-5 & 
61 & 9.88900325e-13 & -9.7738862e-8 \\ 
2 & 5.15493461e-2 & 1.98613310e-1 & 32 & 1.41147446e-8 & 1.22051588e-5 & 62
& -3.89677284e-13 & 7.95035184e-8 \\ 
3 & -1.31953282e-3 & -1.01240820e-1 & 33 & -2.40619550e-9 & -5.65710650e-6 & 
63 & 1.65814880e-14 & -6.28186321e-8 \\ 
4 & -8.60072019e-3 & 4.81900474e-2 & 34 & -3.55320702e-9 & 1.04600755e-6 & 64
& 1.85720723e-13 & 4.80517299e-8 \\ 
5 & 7.39640799e-3 & -1.93134328e-2 & 35 & 5.76787390e-9 & 2.01386153e-6 & 65
& -2.69753981e-13 & -3.53567158e-8 \\ 
6 & -4.37611843e-3 & 4.15396693e-3 & 36 & -5.83318261e-9 & -3.87366749e-6 & 
66 & 2.79353530e-13 & 2.47339511e-8 \\ 
7 & 2.04084356e-3 & 3.18597066e-3 & 37 & 4.87035858e-9 & 4.83802404e-6 & 67
& -2.47807756e-13 & -1.60789691e-8 \\ 
8 & -6.90933604e-4 & -6.14922182e-3 & 38 & -3.58335063e-9 & -5.16109176e-6 & 
68 & 1.98336148e-13 & 9.22096259e-9 \\ 
9 & 5.84170495e-5 & 6.76477185e-3 & 39 & 2.35943870e-9 & 5.04782085e-6 & 69
& -1.45711327e-13 & -3.95235488e-9 \\ 
10 & 1.63910628e-4 & -6.21966082e-3 & 40 & -1.37125048e-9 & -4.65820470e-6 & 
70 & 9.82535981e-14 & 5.08065998e-11 \\ 
11 & -1.91945399e-4 & 5.19268768e-3 & 41 & 6.61230664e-10 & 4.11305389e-6 & 
71 & -5.97484225e-14 & 2.70503277e-9 \\ 
12 & 1.48782636e-4 & -4.05581180e-3 & 42 & -2.03263437e-10 & -3.50028777e-6
& 72 & 3.10645115e-14 & -4.52482301e-9 \\ 
13 & -9.36092580e-5 & 2.99780934e-3 & 43 & -5.68849693e-11 & 2.88110292e-6 & 
73 & -1.13976756e-14 & 5.59980586e-9 \\ 
14 & 4.90747770e-5 & -2.10062357e-3 & 44 & 1.77842259e-10 & -2.29563951e-6 & 
74 & -8.48101535e-16 & -6.09912248e-9 \\ 
15 & -1.99459110e-5 & 1.38633915e-3 & 45 & -2.10698686e-10 & 1.76795021e-6 & 
75 & 7.48776125e-15 & 6.16839162e-9 \\ 
16 & 3.84298932e-6 & -8.45731343e-4 & 46 & 1.94589987e-10 & -1.31020029e-6 & 
76 & -1.02207808e-14 & -5.92990269e-9 \\ 
17 & 3.40159639e-6 & 4.55235540e-4 & 47 & -1.56546402e-10 & 9.26107603e-7 & 
77 & 1.04671090e-14 & 5.48390459e-9 \\ 
18 & -5.50975958e-6 & -1.86675466e-4 & 48 & 1.13383082e-10 & -6.13679173e-7
& 78 & -9.31507219e-15 & -4.91058251e-9 \\ 
19 & 5.11609860e-6 & 1.25168912e-5 & 49 & -7.42527288e-11 & 3.67326262e-7 & 
79 & 7.53574920e-15 & 4.27241069e-9 \\ 
20 & -3.81723065e-6 & 9.15879102e-5 & 50 & 4.31026576e-11 & -1.79450077e-7 & 
80 & -5.63104283e-15 & -3.61664994e-9 \\ 
21 & 2.44320933e-6 & -1.45844751e-4 & 51 & -2.06930189e-11 & 4.15901346e-8 & 
81 & 3.89365101e-15 & 2.97782400e-9 \\ 
22 & -1.33960606e-6 & 1.66248124e-4 & 52 & 6.08330663e-12 & 5.47786577e-8 & 
82 & -2.46576590e-15 & -2.38006211e-9 \\ 
23 & 5.85690505e-7 & -1.64964449e-4 & 53 & 2.37262592e-12 & -1.17723545e-7 & 
83 & 1.38959352e-15 & 1.83923593e-9 \\ 
24 & -1.38111028e-7 & 1.50908098e-4 & 54 & -6.42613977e-12 & 1.54540208e-7 & 
84 & -6.46998338e-16 & -1.36485050e-9 \\ 
25 & -8.61375534e-8 & -1.30360810e-4 & 55 & 7.61941642e-12 & -1.71592877e-7
& 85 & 1.88167674e-16 & 9.61671826e-10 \\ 
26 & 1.68419863e-7 & 1.07554362e-4 & 56 & -7.16001161e-12 & 1.74263724e-7 & 
86 & 4.94043268e-17 & -6.3109027e-10 \\ 
27 & -1.72340268e-7 & -8.51788047e-5 & 57 & 5.90792130e-12 & -1.66976323e-7
& 87 & -1.29706521e-16 & 3.72229392e-10 \\ 
28 & 1.40973391e-7 & 6.48036387e-5 & 58 & -4.42052222e-12 & 1.53266044e-7 & 
88 & 1.12752092e-16 & -1.82817187e-10 \\ 
29 & -9.99911783e-8 & -4.72133771e-5 & 59 & 3.01971648e-12 & -1.35877067e-7
& 89 & -5.29952660e-17 & 5.98398862e-11 \\ \hline\hline
\end{tabular}}  
\label{tab7} } 
 \end{center} 
\end{table} 

   Fig. 1 shows the ground-state wave functions  
$\psi _{0}\left( \lambda ;x\right) $ of
 the anharmonic oscillator, calculated by the above
method, for the coupling constants  $\lambda =1$, $\lambda
=10$, and $\lambda =100$. For comparison, the wave function of the harmonic
oscillator's ground state, $\varphi _{0}\left( x\right) $, is also shown in
Fig. 1. It is obvious and quite understandable that the wave function $\psi
_{0}\left( \lambda ;x\right) $ exhibits a progressively higher, narrower,
and sharper profile with the growth of $\lambda $, mirroring the
increasingly narrow and sharp rise in the perturbation potential $U\left(
x\right) =\lambda x^{4}$ as $\lambda $ increases. Additionally, 
as evident from Fig. 1, the system's wave function decreases very
rapidly with increasing $x$, and this decline becomes even more rapid and
pronounced as $\lambda $ grows.  This can
easily be seen both from the behavior of the perturbation potential as $%
\lambda $ increases and from the nature of the asymptotics (4) of the
anharmonic oscillator's wave function. 
     \begin{figure}[h]
         \centerline{\includegraphics[width=\textwidth]{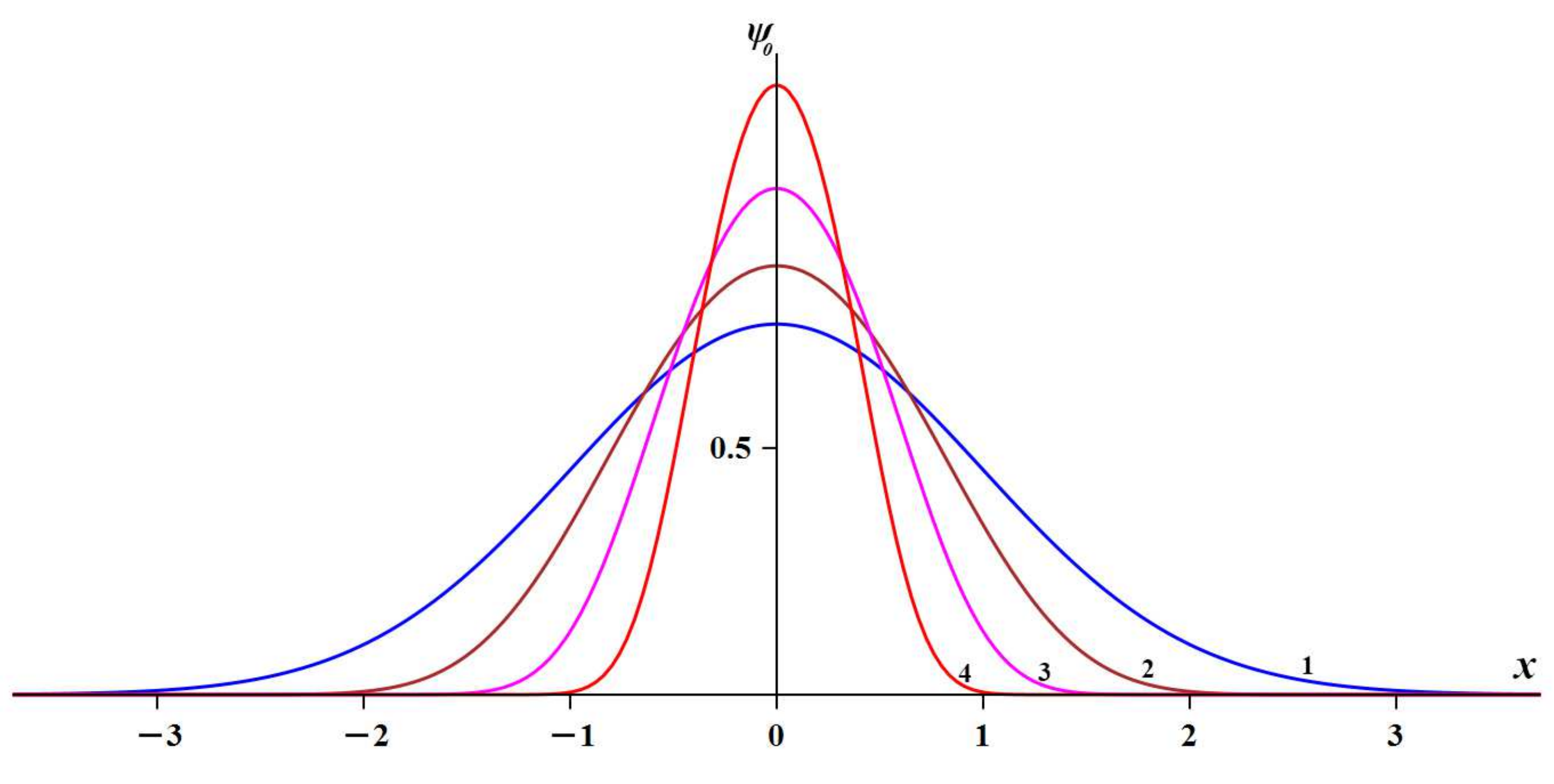}} 
     \caption{The harmonic oscillator ground-state wave function, 
   $\varphi _{0}\left( x\right) $ (blue curve 1), shown along with the ground
   state wave functions of the quartic anharmonic oscillator: $\psi _{0}\left(
   \lambda =1;x\right) $ at a coupling constant of $\lambda =1$ (brown curve
   2), $\psi _{0}\left( \lambda =10;x\right) $ at $\lambda =10$ (lilac curve
   3), and $\psi _{0}\left( \lambda =100;x\right) $ at $\lambda =100$ (red
   curve 4). The last three wave functions were computed and plotted according
   to the expansion in the oscillator basis, utilizing the calculated expansion
   coefficients $c_{2n}$. \label{f1}}
     \end{figure}  

    Fig. 2 presents the wave functions 
$\psi _{1}\left( \lambda ;x\right) $ of the first excited state of the
anharmonic oscillator, calculated in a similar manner, for the
coupling constant values $\lambda =1$, $\lambda =10$, and $\lambda =100$.
 For comparison, Fig. 2 also includes the wave function of the first excited
state of the harmonic oscillator, $\varphi _{1}\left( x\right) $. In general, 
the overall shape of the first excited state wave functions remains the same
with the growth of the coupling constant, as expected.    
   Thus, the presented method offers  a very convenient and accurate
way to calculate both the wave functions of the system's states and its
energy levels for any value of the coupling constant $\lambda $. However,
it is clear that as $\lambda $ increases, especially in the region of strong
coupling, the convergence of expansions naturally deteriorates. 
Therefore, achieving a specified level of accuracy requires incorporating 
an increasing number of basis functions into the expansion.  
In the next
section, however, we will explore a method to substantially improve the convergence
of expansions in oscillator function bases.
  \begin{figure}[h]
       \centerline{\includegraphics[width=\textwidth]{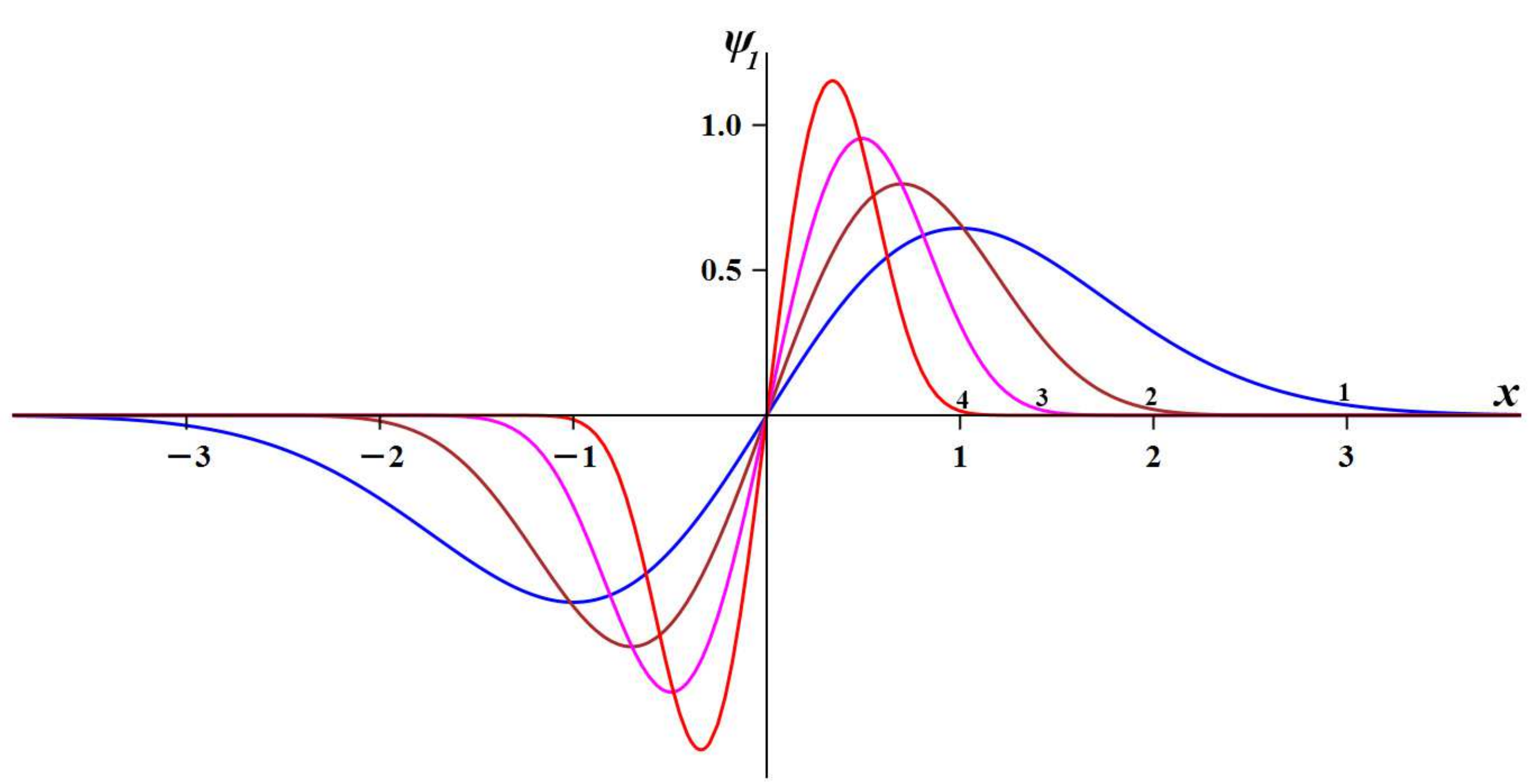}} 
  \caption{The first excited state wave function of the harmonic oscillator, 
$\varphi _{1}\left( x\right) $ (blue curve 1), shown along with the first
excited state wave functions of the  anharmonic oscillator: $\psi
_{1}\left( \lambda =1;x\right) $ at a coupling constant of $\lambda =1$
(brown curve 2), $\psi _{1}\left( \lambda =10;x\right) $ at $\lambda =10$
(lilac curve 3), and $\psi _{1}\left( \lambda =100;x\right) $ at $\lambda
=100$ (red curve 4). The last three wave functions were 
computed and plotted using the oscillator basis expansion 
with the calculated expansion coefficients
 $c_{2n+1}$. \label{f2}}
  \end{figure}

\begin{center}
{\textbf{ 6. Acceleration of Expansion Convergence via Refinement of the 
Zeroth Approximation and Variation of the Harmonic Oscillator Basis Frequency }}  
\end{center}
{\noindent
Let us now address the exceptionally important issue of significantly
 improving the system description in the crucial non-perturbative 
region of strong coupling by accelerating the convergence 
of expansions in the oscillator basis.
The calculations of
energy eigenvalues and corresponding eigenfunctions of the anharmonic
oscillator, according to the method described above, 
clearly demonstrate the deterioration of convergence of expansions 
 as the coupling constant $\lambda $ increases. This
fact is naturally  explained by the 
increasing \textquotedblleft strength\textquotedblright\ of the perturbation
potential $U\left( x\right) =\lambda x^{4}$ with the growth of $\lambda $,
and its correspondingly larger deviation from the harmonic oscillator
potential $\frac{1}{2}x^{2}$, which forms part of the zeroth-approximation
Hamiltonian $H^{\left( 0\right) }=\frac{1}{2}\left( p^{2}+x^{2}\right) $,
whose eigenfunctions serve as the basis for the expansion. 
On the one hand, this 
convergence issue is manageable thanks to   
 modern computing power and efficient matrix diagonalization 
methods.
However, on the other hand, it would
be highly  desirable and convenient to accelerate the convergence of the
employed expansions, especially in the strong coupling region.
Doing so would allow computational results to be obtained more 
quickly and reliably, while also providing a much clearer and 
more comprehensive interpretation.}  
  
    Therefore, presenting approximate solutions and computational results
using a small number ($N\lesssim 10$) of basis functions for expansion,
instead of hundreds or thousands, is very useful and illustrative,
effectively providing results in a compact  analytical form, which is  
exceptionally productive and valuable. 
Moreover, analytical solutions not only facilitate the assessment of the impact 
of different potential parameters but also offer deeper insight 
into how their variations affect the properties of the system under study.
Furthermore, it is important to emphasize  that the calculation of 
wave functions is generally significantly less accurate than that of 
energy eigenvalues. Consequently, achieving much faster convergence of 
expansions can be highly important and beneficial 
for the accurate computation of wave functions.
This accelerated convergence is also crucial for high-precision calculations
involving a large number of significant digits.
Additionally, it may prove extremely important and convenient when
employing the oscillator basis expansion for solving more
computationally complex problems --- such as calculating
the characteristics of more intricate oscillator models
with stronger or more  singular perturbation potentials.
All these factors make the development of methods for accelerating
convergence in the oscillator basis highly relevant and desirable.

Thus, in light of the above, it is extremely important that the presented method
of expansion in the oscillator basis offers such an exceptionally convenient and  
effective opportunity for significant acceleration of 
the convergence of the employed series 
and, overall, a considerable enhancement of the method's efficiency through a more 
effective way of choosing the frequency of the oscillator basis used for expansion. 
It should be noted that methods of renormalization and/or variation of 
the oscillator frequency aimed at obtaining convergent expansions 
and enhancing efficiency have historical roots.
These approaches trace back to considerations of the oscillator 
in classical mechanics, starting
with the works of A. Lindstedt and 
H. Poincar\'{e} --- see, in this context, the Lindstedt--Poincar\'{e} method [61, 62]. 
Additionally, in this context, it is worth mentioning  
studies in which some authors explored ways to 
 improve the convergence of expansions in the
oscillator basis, using simple algebraic models as an example, by
varying the oscillator length, which is directly related to the 
frequency [63--65]. 
Moreover, the use of the oscillator basis with its frequency optimized 
in various ways has been considerably discussed 
within the framework of the variational 
method --- see, e.g., [66--69] and references therein. 
Thus, the method of varying the frequency of the oscillator basis 
that we propose and further utilize is quite natural and well justified. 
Ultimately, it leads to a substantial acceleration of the convergence of 
oscillator expansions and yields very good results.  
   
    Consequently, for further consideration, we will investigate the Hamiltonian of
the quartic anharmonic oscillator with an arbitrary frequency $\omega $: 
\begin{equation}
H=\frac{1}{2}\left( p^{2}+\omega ^{2}x^{2}\right) +\lambda x^{4}~.  \tag{28}
\end{equation}%
Here, we adopt a unit system where the oscillator's mass and the
reduced Planck constant are still set to unity: $m=1$, $\hbar =1$.
In this case, the oscillator frequency $\omega $ is our sole independent
dimensional parameter, so the dimensions of all other physical
characteristics of the oscillator are expressed through the dimension of
frequency, as follows: $%
\left[ H\right] =\left[ E\right] =\left[ \omega \right] $, $\left[ p\right] =%
\left[ \omega \right] ^{1/2}$, $\left[ x\right] =\left[ \omega \right]
^{-1/2}$, $\left[ \lambda \right] =\left[ \omega \right] ^{3}$. It is well
known that frequency is typically measured in hertz --- $1$Hz$~=1s^{-1}$.
However, since in our case ($\hbar =1$) the dimension of frequency coincides
with that of energy, 
in various quantum mechanical contexts
 frequency and energy can be measured in the corresponding units
accepted in the respective field. For example, in nuclear physics energy is most
often measured in mega-electronvolts (MeV), while in atomic physics it is
measured in electronvolts (eV) or in Hartree energies ($1$E$_{\text{h}%
}=27.2113845$eV). 
Going forward, we will omit the units of measurement for the 
corresponding quantities, since our ultimate goal remains the calculation of
characteristics of Hamiltonian (1) with unit frequency, i.e., in the end,
we will be calculating all quantities for the previously considered case of $%
\omega =1$, and ultimately transitioning to this case.

   For small values of the coupling constant $\lambda $, which are typically
considered first, it was quite natural, as discussed earlier, to take the
Hamiltonian corresponding to the zero value of the coupling constant $%
\lambda =0$ as the zeroth-approximation Hamiltonian: 
\begin{equation}
H^{\left( 0\right) }=H\left( \lambda =0\right) =\frac{1}{2}\left(
p^{2}+\omega ^{2}x^{2}\right) ~.  \tag{29}
\end{equation}%
Accordingly, in this case, the complete set of eigenfunctions of the
Hamiltonian $H^{\left( 0\right) }$ was utilized as the basis for the
expansion of the system's wave function. However, at high values of the
parameter $\lambda $, especially as it increases in the strong
coupling region, the zeroth approximation (29) becomes increasingly poor and
less adequate, which in turn leads to the deterioration of convergence of the
computed physical quantities characterizing the anharmonic oscillator. 
    
    Nevertheless,  
it is possible to significantly improve the situation 
and fundamentally enhance the zeroth approximation in our problem by
choosing the frequency of the zeroth-approximation Hamiltonian 
more effectively and rationally.  
Specifically, let us consider a certain harmonic oscillator Hamiltonian  
$\mathscr{H}^{\left( 0\right) }\left( \omega 
_{0}\right)$ 
as the zeroth-approximation Hamiltonian, whose eigenfunctions 
will be used for the expansion. This newly introduced Hamiltonian corresponds to some 
effective frequency $\omega _{0}$, which generally differs from the 
previously mentioned $\omega $ and will be treated 
 as a certain free, variable, fitting parameter. This  
zeroth-approximation Hamiltonian is expressed as%
\begin{equation}
\mathscr{H}^{\left( 0\right) }=\mathscr{H}^{\left( 0\right) }\left( \omega
_{0}\right) =\frac{1}{2}\left( p^{2}+\omega _{0}^{2}x^{2}\right) ~,  \tag{30}
\end{equation}%
and through an effective and reasonable choice of the frequency $\omega _{0}$
(by increasing it), it can be made a much better and closer-to-optimal zeroth
approximation, significantly closer to the initial total Hamiltonian (28)
than Hamiltonian (29), $H^{\left( 0\right) }$. The total Hamiltonian of the
system $H$ can now be written as: 
\begin{equation}
H=\mathscr{H}^{\left( 0\right) }+W~,  \tag{31}
\end{equation}%
where the perturbation operator $W$ is given by the expression: 
\begin{equation}
W=\frac{\omega ^{2}-\omega _{0}^{2}}{2}x^{2}+\lambda x^{4}~.  \tag{32}
\end{equation} 

   Expressions (30)--(32) clearly demonstrate the advantage of choosing 
the zeroth-approximation Hamiltonian in the form of (30) with 
an appropriately selected effective frequency $\omega _{0}$.   
Namely,  as the parameter $\lambda $ increases, 
particularly at high fixed values of $%
\lambda $ from the strong coupling region, we can make the potential of the
zeroth-approximation Hamiltonian $\frac{1}{2}\omega _{0}^{2}x^{2}$ significantly 
\textquotedblleft closer\textquotedblright\ to the perturbation potential $%
U\left( x\right) =\lambda x^{4}$ by increasing the parameter $\omega _{0}$,
thereby substantially improving 
 the quality of this zeroth approximation.
Equivalently, for a fixed $\lambda $, the
perturbation potential $W$ can be effectively and substantially
\textquotedblleft diminished\textquotedblright\ by increasing the parameter $%
\omega _{0}$, since the corresponding term in (32) appears with a negative
sign. The above is illustrated in Fig. 3, which shows the potential of the
harmonic oscillator in the case of the usual zeroth approximation ($\omega
_{0}=1$) and in the case of the improved zeroth approximation ($\omega
_{0}=16$), compared to the perturbation potential $U\left( x\right) =\lambda
x^{4}$ at $\lambda =100$, chosen as an example. In this case, the frequency
parameter of the improved zeroth approximation, $\omega _{0}=16$,  
was determined based on
 our further calculations (see Table 9) 
as  the best choice
 for computations with the perturbation potential $U\left( x\right)
=100x^{4}$. Thus, a more effective choice of the frequency parameter $\omega
_{0}$ in the zeroth approximation (30), specifically by increasing it for
cases of higher values of the coupling constant $\lambda $, can greatly
enhance the quality and adequacy of the zeroth approximation (30). 
%
  \begin{figure}[h]
       \centerline{\includegraphics[width=\textwidth]{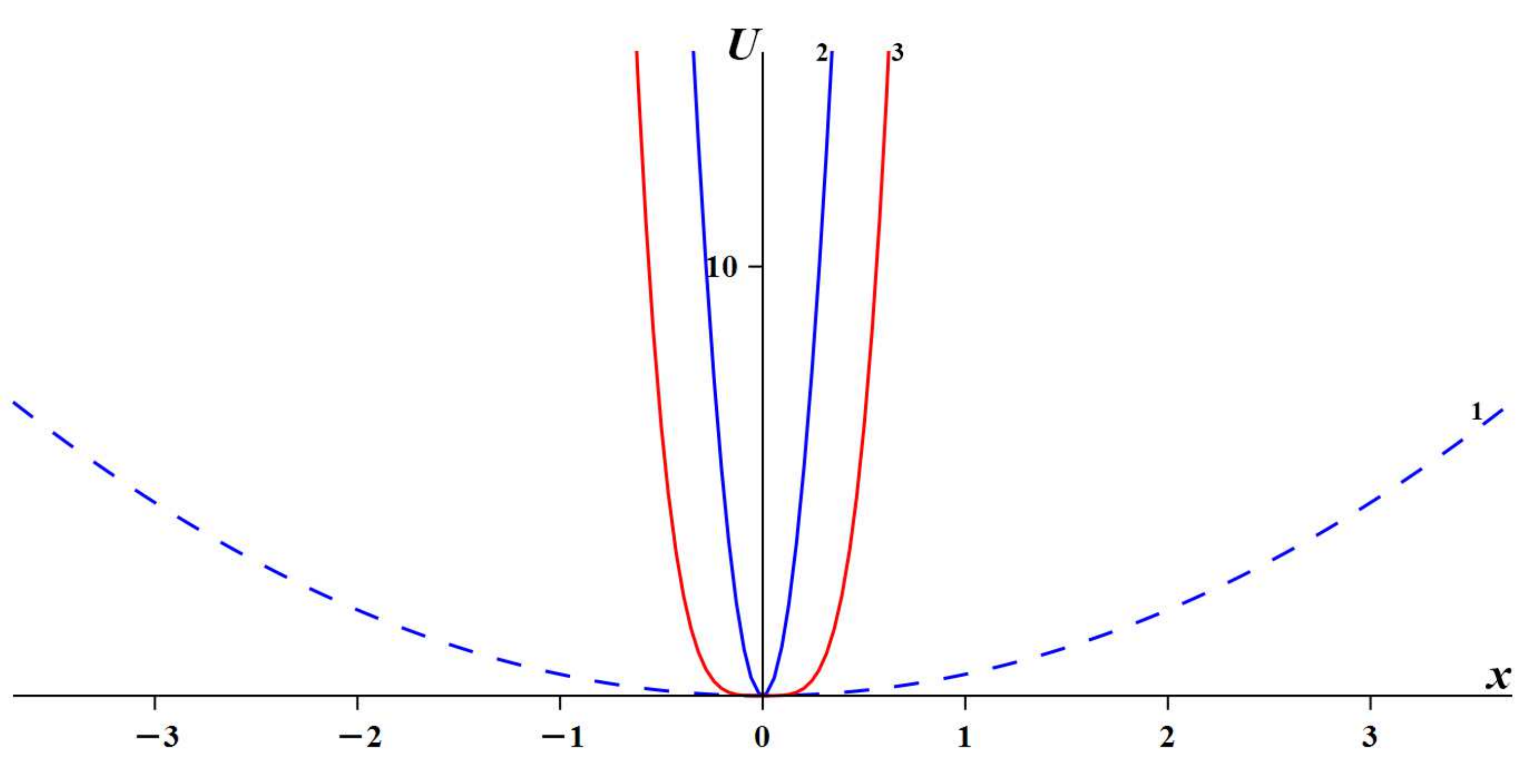}} 
  \caption{The harmonic oscillator potential $\frac{1}{2}\omega _{0}^{2}x^{2}$
  at a frequency of $\omega _{0}=1$ (dashed blue curve 1), corresponding to
  the usual zeroth approximation, and at a frequency of $\omega _{0}=16$
  (solid blue curve 2), corresponding to the improved zeroth approximation, as
  well as the perturbation potential $U\left( x\right) =\lambda x^{4}$ at a
  coupling constant of $\lambda =100$ (red curve 3). \label{f3}}
  \end{figure}

   This phenomenon  is also evident from Figs. 1 and 2, which effectively
demonstrate, through the example of the corresponding wave functions, the
gradual deterioration of the quality of the usual zeroth
approximation as the parameter $\lambda $ increases. It is confirmed by
the growing discrepancy between the true wave function of the system and the
oscillator wave function of the zeroth approximation as $\lambda $
increases. 
However, by increasing the frequency $\omega _{0}$ of the basis wave functions 
of the Hamiltonian (30), they become more localized near zero and sharper, 
bringing them closer to the system's true wave function.
At the same time, as $\omega _{0}$ increases, the improvement of the zeroth 
approximation (30) cannot continue indefinitely and eventually reaches a limit.  
Beyond 
a certain point, the quality of approximation (30) will begin 
to deteriorate, due to an increasing deviation of the potential 
and the main wave function of the zeroth approximation 
 from the true values in the opposite direction.
In other words, there must inevitably  exist an optimal value of the
parameter $\omega _{0}$, at which the system's description and the
convergence of the corresponding expansions are at their best. Thus,
overall, selecting the  optimal zeroth approximation (30), with the 
best choice of the frequency $\omega _{0}$, is based on clear and
transparent mathematical and physical reasoning.
This is justified both by
\textquotedblleft effectively\textquotedblright\ reducing the perturbation
potential $W$ in this approach and by ensuring that the wave
function of the zeroth approximation becomes increasingly closer to the true
wave function, thereby enhancing the adequacy of the zeroth approximation.
It should also be noted that the discussions in the previous sections
clearly correspond to the special limiting case of $\omega _{0}=\omega =1$. 
Thus, the present consideration is a natural generalization 
of the previously discussed formalism. 

   To specifically develop and implement  the above-described plan for improving
the system's description and accelerating the convergence of expansions in
the oscillator basis, we will now employ the well-known form [57, 58] of
the eigenfunctions $\varphi _{n}\left( \omega _{0};x\right) $ corresponding
to the Hamiltonian of a general harmonic oscillator $\mathscr{H}^{\left(
0\right) }\left( \omega _{0}\right) $ with some arbitrary frequency $\omega
_{0}$: 
\begin{equation}
\varphi _{n}\left( \omega _{0};x\right) =\frac{\sqrt[4]{\omega _{0}/\pi }}{%
\sqrt{2^{n}n!}}e^{-\omega _{0}\frac{x^{2}}{2}}H_{n}\left( \sqrt{\omega _{0}}%
x\right) ,~n=0,~1,~2,\ldots .  \tag{33}
\end{equation}%
The eigenfunctions $\left\{ \varphi _{n}\left( \omega _{0};x\right) \right\}
_{n=0}^{\infty }$ of the Hamiltonian $\mathscr{H}^{\left( 0\right) }\left(
\omega _{0}\right) $ correspond to its energy eigenvalues%
\begin{equation}
\mathscr{E}^{\left( 0\right) }_n=\omega _{0}\left( n+\frac{1}{2}\right)
~,~~~n=0,~1,~2,\ldots  \tag{34}
\end{equation}%
and form a complete set, or basis, in the space of square-integrable
functions $L^{2}\left( 
\mathbb{R}
\right) $. 

To distinguish it from the earlier considerations and ensure 
a more precise and clear identification, we will 
henceforth refer to the basis $\left\{ \varphi _{n}\left( \omega
_{0};x\right) \right\} _{n=0}^{\infty }$, corresponding to the Hamiltonian $%
\mathscr{H}^{\left( 0\right) }\left( \omega _{0}\right) $, as the
modified oscillator basis or the optimizing basis. In order to shift to
more convenient dimensionless variables when considering the Hamiltonian $%
\mathscr{H}^{\left( 0\right) }\left( \omega _{0}\right) $ and its
eigenfunctions $\varphi _{n}\left( \omega _{0};x\right) $, it is common to
introduce an additional characteristic parameter with the dimension of
length, usually denoted as $x_{0}$ or $b_{0}$, known as the
oscillator length. In our case ($m=1$, $\hbar =1$), the oscillator length $%
x_{0}$ is defined by the relation%
\begin{equation}
x_{0}=\frac{1}{\sqrt{\omega _{0}}}~,  \tag{35}
\end{equation}%
while in the general case, it is defined by the well-known relation $x_{0}=%
\sqrt{\hbar /m\omega _{0}}$. Then, the dimensionless coordinate $\xi $ is
defined as%
\begin{equation}
\xi =\frac{x}{x_{0}}=\sqrt{\omega _{0}}x~,  \tag{36}
\end{equation}%
and the eigenfunctions $\varphi _{n}\left( \omega _{0};x\right) $ can be
expressed as:%
\begin{equation}
\varphi _{n}\left( \omega _{0};x\right) =\frac{1}{\sqrt{x_{0}}}\varphi
_{n}\left( \xi \right) ~,  \tag{37}
\end{equation}%
where the functions $\varphi _{n}\left( \xi \right) $ are given by
expression (10), with the independent variable there understood as
dimensionless. 
  
     It should be noted that, in terms of the oscillator length $%
x_{0}$, an additional physical interpretation or justification for the
improvement in the efficiency of the zeroth approximation (30) is possible
with the appropriate choice of the parameter $\omega _{0}$. Specifically, as
the characteristic oscillator length $x_{0}$ decreases, corresponding to an
increase in the frequency $\omega _{0}$, the zeroth-approximation wave
function becomes increasingly localized near the origin, and 
its shape increasingly resembles the true wave function 
 of the system at large
values of the coupling constant $\lambda $, i.e., in the case of a more
rapid and steep increase in the perturbation potential $U\left( x\right)
=\lambda x^{4}$.

   The square-integrable wave function $\psi \left( x\right) $ of our system
can be expanded, as in the previous discussion,
  into a convergent series over
the complete set of modified basis functions (33) of the Hamiltonian $%
\mathscr{H}^{\left( 0\right) }\left( \omega _{0}\right) $ of 
the $\omega_{0}$-related 
harmonic oscillator:{\ 
\begin{equation}
\psi \left( x\right) =\overset{\infty }{\underset{n=0}{\dsum }}c_{n}\varphi
_{n}\left( \omega _{0};x\right) ~,  \tag{38}
\end{equation}%
}and, analogously to Section 2, this leads to an infinite homogeneous
system of linear algebraic equations of the form (17) for the coefficients $%
c_{n}$. The complete set of these coefficients $\left\{ c_{n}\right\}
_{n=0}^{\infty }=\left\{ \left\langle \varphi _{n}\left( \omega _{0}\right)
\mid \psi \right\rangle \right\} _{n=0}^{\infty }$ represents the wave
function of our system in the representation of the harmonic oscillator
defined by the Hamiltonian $\mathscr{H}^{\left( 0\right) }\left( \omega
_{0}\right) $, and these coefficients obviously depend on the frequency $%
\omega _{0}$ as a parameter. However, now the matrix elements $H_{mn}$ of
the total Hamiltonian $H$ of our system will be determined by a different
expression, according to its decomposition $H=\mathscr{H}^{\left(
0\right) }+W$ into the unperturbed part $\mathscr{H}^{\left( 0\right) }$ and
the perturbation operator $W$: 
\begin{equation}
H_{mn}=\left\langle \varphi _{m}\left( \omega _{0}\right) \mid H\mid \varphi
_{n}\left( \omega _{0}\right) \right\rangle =\mathscr{H}_{mn}^{\left(
0\right) }+W_{mn}~.  \tag{39}
\end{equation}

In this case, the matrix elements of the zeroth-approximation Hamiltonian $%
\mathscr{H}^{\left( 0\right) }$ (see (30)), i.e., the Hamiltonian of the
$\omega_{0}$-related harmonic oscillator, are diagonal in the eigenbasis and, in
accordance with (34), take the form: 
\begin{equation}
\mathscr{H}_{mn}^{\left( 0\right) }=\omega _{0}\left( n+\frac{1}{2}\right)
\delta _{mn}~.  \tag{40}
\end{equation}%
Meanwhile, the matrix elements of the perturbation 
operator $W$, following from (32), are given by:  
\begin{equation}
W_{mn}=\frac{\omega ^{2}-\omega _{0}^{2}}{2}~\left\langle x^{2}\right\rangle
_{mn}+\lambda \left\langle x^{4}\right\rangle _{mn}~.  \tag{41}
\end{equation}%
 For the matrix elements $\left\langle x^{2}\right\rangle _{mn}$ and $%
\left\langle x^{4}\right\rangle _{mn}$, it is straightforward to obtain,
using the relation $x=x_{0}\xi $ and expression (37), their
representation in terms of the dimensionless matrix elements of the
dimensionless coordinate $\xi $: 
\begin{equation}
\left\langle x^{2}\right\rangle _{mn}=x_{0}^{2}~\left\langle \xi
^{2}\right\rangle _{mn}=\frac{1}{\omega _{0}}\left\langle \xi
^{2}\right\rangle _{mn}~,  \tag{42}
\end{equation}%
\begin{equation}
\left\langle x^{4}\right\rangle _{mn}=x_{0}^{4}~\left\langle \xi
^{4}\right\rangle _{mn}=\frac{1}{\omega _{0}^{2}}\left\langle \xi
^{4}\right\rangle _{mn}~.  \tag{43}
\end{equation}%
Here, the matrix elements of $\xi ^{2}$ and $\xi ^{4}$, which are powers of
the dimensionless coordinate, are taken with respect to the dimensionless
basis functions $\varphi _{n}\left( \xi \right) $, unlike the matrix
elements of the dimensional coordinate $x$. The explicit form of the matrix
elements $\left\langle \xi ^{4}\right\rangle _{mn}$ is still given by
formula (20), where $x$ should be understood as the dimensionless variable.
Additionally, the explicit form of the matrix elements $\left\langle \xi
^{2}\right\rangle _{mn}$ follows directly from the recurrence
relations (13) and can be written as:   
\begin{equation*}
\left\langle \xi ^{2}\right\rangle _{mn}=\left\langle \varphi _{m}\left( \xi
\right) \mid \xi ^{2}\mid \varphi _{n}\left( \xi \right) \right\rangle =
\end{equation*}%
\begin{equation}
=\frac{1}{2}\sqrt{n\left( n-1\right) }\delta _{m,n-2}+\left( n+\frac{1}{2}%
\right) \delta _{mn}+\frac{1}{2}\sqrt{\left( n+1\right) \left( n+2\right) }%
\delta _{m,n+2}~.  \tag{44}
\end{equation}

   Thus, finally, taking into account (39)--(43), we can write the matrix
elements $H_{mn}$ of the total Hamiltonian of our system as: 
\begin{equation}
H_{mn}=\omega _{0}\left( n+\frac{1}{2}\right) \delta _{mn}+\frac{\omega
^{2}-\omega _{0}^{2}}{2\omega _{0}}\left\langle \xi ^{2}\right\rangle _{mn}+%
\frac{\lambda }{\omega _{0}^{2}}\left\langle \xi ^{4}\right\rangle
_{mn},~m,n=0,~1,~2,\ldots ~.  \tag{45}
\end{equation}%
If, for a specific numerical solution of the infinite system of linear
algebraic equations of the form (17), we truncate the expansion (38) and,
accordingly, the system (17) to a certain finite number of terms $M$, then
by substituting the obtained matrix elements (45) into (17), we can write
this system of homogeneous linear algebraic equations for determining the
coefficients $c_{n}$ in the following finite form: 
\begin{equation}
\overset{M}{\underset{n=0}{\dsum }}\left[ \omega _{0}\left( n+\frac{1}{2}%
\right) \delta _{mn}+\frac{\omega ^{2}-\omega _{0}^{2}}{2\omega _{0}}%
\left\langle \xi ^{2}\right\rangle _{mn}+\frac{\lambda }{\omega _{0}^{2}}%
\left\langle \xi ^{4}\right\rangle _{mn}\right] c_{n}=Ec_{m},~~m=0,1,\ldots,M~. 
 \tag{46}
\end{equation}%
In the last two formulas (45) and (46), the matrix elements $%
\left\langle \xi ^{2}\right\rangle _{mn}$ and $\left\langle \xi
^{4}\right\rangle _{mn}$ are explicitly given by formulas (44) and
(20), respectively. For the final numerical calculations based on 
system (46), aimed at determining the properties of specific physical
states of the anharmonic oscillator (i.e., calculating the energy $E$
eigenvalues and the coefficients $c_{n}$), we must  
also account for the parity of these states, as described 
earlier.  
In particular, for the ground state  
and other even states, the expansion (38) in the optimizing
oscillator basis will contain only even basis functions ($c_{2j+1}=0$
for even states, $M=2N$), while for odd states, the expansion (38) will
contain only odd basis functions ($c_{2j}=0$ for odd states, $M=2N+1$).

   Then, in Table 8, for various values of ${\lambda }$, the results of
calculations for the eigenvalues of the ground state energy {$E_{0}\left( {%
\lambda }\right) $} of the quartic anharmonic oscillator (Hamiltonian (28)
with the frequency $\omega =1$) are presented with standard precision of
eight significant digits after the decimal. These values are calculated in  
various orders of the expansion (38) of the system's wave function over the
modified optimizing oscillator basis with a specifically chosen frequency 
$\omega _{0}$. Calculations were performed using 
the previously described optimized method, 
i.e., energy values were calculated as the eigenvalues
of the equation system (46) with a suitably fitted  
value of $\omega _{0}$. In this
case, the values of the coupling constant ${\lambda }$ in Table 8 are selected 
in the range of intermediate and strong coupling 
{$1\lesssim {\lambda } \lesssim 50$}.  
\begin{table}[h] 
\captionsetup{width=0.94\textwidth} 
{\boldmath{ 
\caption{\footnotesize \textbf{The ground state energy values $E_{0}$
of the quartic anharmonic oscillator with frequency $\omega =1$,
calculated in various orders $N$ of the expansion (38) in the modified
optimizing oscillator basis  
$\left\{ \varphi _{n}\left( \omega_{0};x\right) \right\} $  
for certain values of the basis frequency  
$\omega _{0}$. The values of the variable fitting frequency $\omega _{0}$ 
of the harmonic oscillator basis, employed for the expansion, were
empirically chosen to ensure the best convergence rate. The 
coupling constant $\lambda $ values are taken from the intermediate and strong
coupling region $1\lesssim {\lambda }\lesssim 50$.}}  
}} 
\vspace*{-1.5em} 
 \begin{center} 
{ 
\setlength{\tabcolsep}{5pt} 
{\footnotesize 
\begin{tabular}{@{}cccccccc@{}}
\hline\hline 
$N$ & $\lambda =1,$ & $\lambda =2,$ & $\lambda =5,$ & $\lambda =10,$ & $%
\lambda =20,$ & $\lambda =25,$ & $\lambda =50,$ \\ 
& $\omega _{0}=4.5$ & $\omega _{0}=4.5$ & $\omega _{0}=4.5$ & $\omega
_{0}=7.5$ & $\omega _{0}=9.0$ & $\omega _{0}=9.0$ & $\omega _{0}=12.0$ \\ 
\hline
1 & 0.85896477 & 0.96685111 & 1.22773634 & 1.53243233 & 1.88704830 & 
2.00960516 & 2.52418732 \\ 
2 & 0.80775832 & 0.95161545 & 1.22534297 & 1.50508881 & 1.86617258 & 
2.00337901 & 2.50059745 \\ 
3 & 0.80384895 & 0.95161476 & 1.22466262 & 1.50507306 & 1.86580169 & 
2.00203457 & 2.49983826 \\ 
4 & 0.80377138 & 0.95156872 & 1.22458924 & 1.50497296 & 1.86569885 & 
2.00200816 & 2.49971522 \\ 
5 & 0.80377090 & 0.95156870 & 1.22458854 & 1.50497295 & 1.86569634 & 
2.00199650 & 2.49970938 \\ 
6 & 0.80377065 & 0.95156847 & 1.22458712 & 1.50497241 & 1.86569585 & 
2.00199650 & 2.49970887 \\ 
7 & 0.80377065 & 0.95156847 & 1.22458704 & 1.50497241 & 1.86569580 & 
2.00199639 & 2.49970877 \\ 
8 & 0.80377065 & 0.95156847 & 1.22458704 & 1.50497241 & 1.86569580 & 
2.00199638 & 2.49970877 \\ 
9 & 0.80377065 & 0.95156847 & 1.22458704 & 1.50497241 & 1.86569580 & 
2.00199638 & 2.49970877 \\ 
10 & 0.80377065 & 0.95156847 & 1.22458704 & 1.50497241 & 1.86569580 & 
2.00199638 & 2.49970877 \\ \hline
Ex. & 0.80377065 & 0.95156847 & 1.22458704 & 1.50497241 & 1.86569580 & 
2.00199638 & 2.49970877 \\ \hline\hline
\end{tabular} }  
\label{tab8} }
 \end{center} 
\end{table} 

   Similar results for the ground state energy   
{$E_{0}\left( {%
\lambda }\right) $}  
 at coupling constant values 
from the superstrong coupling region ${\lambda \gtrsim 100}$ are
presented in Table 9. Tables 8 and 9 clearly and distinctly demonstrate the
exceptionally high convergence rate and superior effectiveness of the method
when the variable frequency parameter $\omega _{0}$ is appropriately chosen
for a given fixed value of the coupling constant ${\lambda }$. The results
show that the convergence rate, achieved through variation and optimal
choice of the frequency $\omega _{0}$, does not decrease and remains at the
same very high level across the entire range of the parameter ${\lambda }$.
Thus, the proposed method enables the computation of 
the anharmonic oscillator's 
 energy values with standard precision over the
entire range of ${\lambda }$, 
while employing an extremely small set of $N\lesssim 10$ basis
functions $\varphi _{n}\left( \omega _{0};x\right) $ without increasing
their number even as ${\lambda }$ increases. 
     As a result, practical calculations
across the entire range of variation of the coupling constant ${\lambda }$ are
realistically achievable using a modified optimizing oscillator basis of
very small dimension $N\lesssim 10$, as the usual deterioration in
convergence rate with an increase in the parameter ${\lambda }$ is not
observed in this case. 
Overall, the approach based on expansion over an optimizing
oscillator basis with a variable fitting frequency effectively solves the
problem of strong coupling in this case, as the number of basis functions
required for calculations does not increase 
with the growth of $\lambda $.
Given that in many practical applications the required 
computational accuracy does not exceed two or three significant digits after 
the decimal, Tables 8 and 9 suggest that a modified basis 
of only three to four functions is sufficient for calculations 
in such cases. This demonstrates  that the solution can be effectively represented   
in analytic form under these conditions.
\begin{table}[h] 
\captionsetup{width=0.92\textwidth}
{\boldmath{ 
\caption{\footnotesize \textbf{The ground state energy values $E_{0}$
of the quartic anharmonic oscillator with frequency $\omega =1$,
calculated in various orders $N$ of the expansion (38) in the modified
optimizing oscillator basis 
$\left\{ \varphi _{n}\left( \omega_{0};x\right) \right\} $ 
for certain values of the basis frequency 
$\omega _{0}$. The values of the variable fitting frequency $\omega _{0}$
of the harmonic oscillator basis, employed for the expansion, were
empirically chosen to ensure the best convergence rate. The  
coupling constant $\lambda $ values are taken from the superstrong coupling region 
${\lambda \gtrsim 100}$.}}  
}}  
\vspace*{-1.5em} 
 \begin{center} 
{ 
\setlength{\tabcolsep}{5pt}  
{\footnotesize 
\begin{tabular}{@{}cccccccc@{}}
\hline\hline
$N$ & $\lambda =100,$ & $\lambda =500,$ & $\lambda =1000,$ & $\lambda
=2000,$ & $\lambda =5000,$ & $\lambda =10000,$ & $\lambda =20000,$ \\ 
& $\omega _{0}=16.0$ & $\omega _{0}=17.0$ & $\omega _{0}=18.5$ & $\omega
_{0}=24.5$ & $\omega _{0}=33.0$ & $\omega _{0}=60.5$ & $\omega _{0}=84.0$ \\ 
\hline
1 & 3.19240524 & 5.39822513 & 6.82756134 & 8.58844817 & 11.65172065 & 
14.41269727 & 18.22762509 \\ 
2 & 3.13167746 & 5.32132402 & 6.70885544 & 8.43777773 & 11.44619041 & 
14.41267371 & 18.14967664 \\ 
3 & 3.13162371 & 5.32092453 & 6.69477345 & 8.42843302 & 11.43200148 & 
14.39821432 & 18.13771217 \\ 
4 & 3.13138558 & 5.32009377 & 6.69477316 & 8.42813470 & 11.43169197 & 
14.39810137 & 18.13733759 \\ 
5 & 3.13138551 & 5.31990047 & 6.69435553 & 8.42758389 & 11.43093573 & 
14.39800558 & 18.13723048 \\ 
6 & 3.13138419 & 5.31989777 & 6.69422972 & 8.42750156 & 11.43080967 & 
14.39799568 & 18.13723029 \\ 
7 & 3.13138417 & 5.31989555 & 6.69422250 & 8.42750105 & 11.43080824 & 
14.39799558 & 18.13722911 \\ 
8 & 3.13138417 & 5.31989445 & 6.69422213 & 8.42749926 & 11.43080603 & 
14.39799536 & 18.13722907 \\ 
9 & 3.13138416 & 5.31989436 & 6.69422117 & 8.42749831 & 11.43080465 & 
14.39799534 & 18.13722907 \\ 
10 & 3.13138416 & 5.31989436 & 6.69422088 & 8.42749818 & 11.43080445 & 
14.39799534 & 18.13722907 \\ 
11 & 3.13138416 & 5.31989436 & 6.69422085 & 8.42749818 & 11.43080444 & 
14.39799534 & 18.13722907 \\ 
12 & 3.13138416 & 5.31989436 & 6.69422085 & 8.42749818 & 11.43080444 & 
14.39799534 & 18.13722907 \\ \hline
Ex. & 3.13138416 & 5.31989436 & 6.69422085 & 8.42749818 & 11.43080444 & 
14.39799534 & 18.13722907 \\ \hline\hline
\end{tabular} }  
\label{tab9} }
 \end{center} 
\end{table}
 
   All of this is evidently achievable  in each specific case of the given
parameter ${\lambda }$ solely due to the correct and successful choice of the
frequency $\omega _{0}$ of the modified oscillator basis. It is precisely this 
right choice that ensures exceptionally high quality and accuracy of the zeroth
approximation, i.e., it effectively brings the zeroth approximation, 
specifically, the Hamiltonian $\mathscr{H}^{\left( 0\right) }\left( \omega
_{0}\right) $, substantially closer to the initial Hamiltonian $H$ 
by appropriately fitting the 
 frequency $\omega _{0}$. Moreover, as previously 
 noted, this means that with a fixed ${\lambda }$, it is possible in
this approach to significantly \textquotedblleft effectively
reduce\textquotedblright\ the strength of the perturbation potential $W$
through a corresponding increase in the parameter $\omega _{0}$ --- see
formula (32) and the text following it. The above is clearly and effectively
demonstrated in Tables 8 and 9, particularly in their first few rows,  
from which it is evident that accounting for only the first two or
three terms of the expansion in the modified oscillator basis, with a
suitably selected $\omega _{0}$,  
ensures that the accuracy of the calculated energy values is 
 ${\gtrsim 95\%}$ relative to the true value. This further confirms 
the high quality of the zeroth approximation with a properly chosen 
parameter $\omega _{0}$.   
   
     At the same time, from Tables 8 and 9, 
 it is clear  that, as 
 ${\lambda }$ increases, the necessary
increase in the variable frequency parameter $\omega _{0}$ 
remains quite moderate and does not
exceed  
 two orders of magnitude  compared to the initial value $\omega
_{0}=1$, which  corresponds to the discussions in Sections 2--5. Thus, the
parameter $\omega _{0}$ varies within a relatively narrow range $1\leqslant
\omega _{0}\leqslant 100$ for all calculations in the practically
significant region of ${\lambda }$ variation. 
This fact suggests a fairly convenient and not very labor-intensive 
procedure for numerically selecting an optimal specific value 
of the parameter $\omega _{0}$, especially since the parameter  
$\omega _{0}$ can be fixed in practical calculations with a  
moderate accuracy of $\sim 0.2\div 0.7$.
Such accuracy is generally sufficient to
reliably establish the region of best convergence in terms of $N$, namely,
the number of modified oscillator basis functions used in the expansion.
And overall, choosing the optimal value of the basis frequency $\omega _{0}$%
, which ensures the highest convergence rate for the expansions, is achieved
quite easily and reliably by incrementally increasing this frequency as the
values of the coupling constant ${\lambda }$ increase, for which the
corresponding calculations need to be performed. This is all the more easily
accomplished in specific calculations due to the  increasingly rapid
convergence of the applied expansions and the need to consider only
a small number of optimizing oscillator basis functions.  
     Thus, the proposed
optimized method enables a very convenient, quick, and effective calculation
of the ground state energy of the anharmonic oscillator over the entire
practically significant range of variation of the parameter ${\lambda }$, 
from zero to arbitrarily large values. 
According to this method, it is
also possible to calculate the energy values of the excited states in a
similar manner, as the corresponding eigenvalues of the system of equations
(46), i.e., the eigenvalues of the finite Hamiltonian matrix with elements
(45) in the representation of the modified harmonic oscillator, defined
by the zeroth-approximation Hamiltonian $\mathscr{H}^{\left( 0\right)
}\left( \omega _{0}\right) $. Moreover, the increase in the order $N$ of the
optimizing oscillator basis required for the calculation of excited states
is quite moderate, and the corresponding calculations based on solving the
system of equations (46) present no difficulties --- 
see the subsequent text for more details.

   The empirically chosen values of the variable fitting frequency $\omega _{0}$
for each ${\lambda }$, determined for the optimizing oscillator basis and
presented in Tables 8 and 9, enable  a straightforward description of this
empirical dependence $\omega _{0}\left( {\lambda }\right) $ using a simple
formula. Since the coupling constant ${\lambda }$ varies over a wide range
 in our case, it is practically convenient to investigate the
dependence of the frequency $\omega _{0}$ on the parameter $\mu =\ln \lambda 
$ (strictly speaking, on the parameter 
$\ln (\lambda / \omega^3)$, but in this work, 
we always assume $\omega=1$). We will express this relation 
in the form of a simple linear dependence 
formula with a power correction: 
\begin{equation}
\omega _{0}=a+b\mu +c\mu ^{\alpha }~,~~~\mu =\ln \lambda ~,  \tag{47}
\end{equation}%
where the values of the parameters $a$, $b$, $c$, and $\alpha $ will be
determined by fitting this dependence using the least squares method. The
resulting values of these parameters are as follows: 
\begin{equation}
a=3.47542~,~~~b=1.92476~,~~~c=2.25163\cdot 10^{-7}~,~~~\alpha =8.48258~. 
\tag{48}
\end{equation}%
  \begin{figure}[h]
       \centerline{\includegraphics[width=\textwidth]{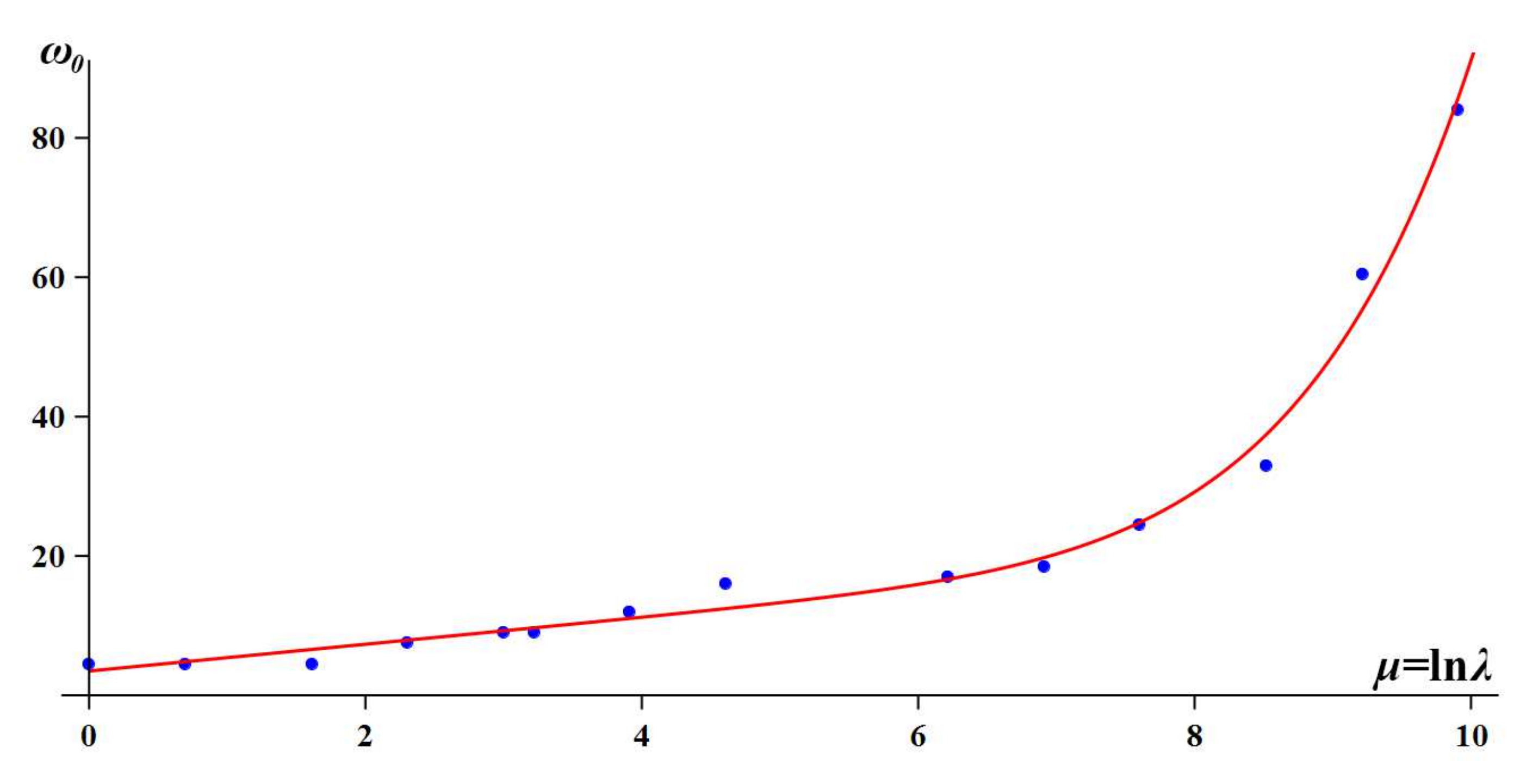}} 
  \caption{Values of the variable fitting frequency $\omega _{0}$ of the
optimizing oscillator basis from Tables 8 and 9 as a function of the
parameter $\mu =\ln \lambda $ (blue points), along with the description of
this dependence by the empirical formula (47) with parameters (48) (red
curve). \label{figadda}}
  \end{figure} 

    Formula (47), with the parameter values (48), describes the empirical
dependence $\omega _{0}\left( \mu \right) $ very well, as illustrated  in
Fig. 4.  It also shows that the observed spread 
in the calculated values of the frequency $\omega
_{0}$ can be refined to ensure 
a smoother behavior  of this phenomenological relation  by further adjusting  
the empirical values of $\omega _{0}$. However, in reality,   
precise calculations and exact determinations of the $\omega _{0}$ 
values are not strictly  necessary, as the high convergence rate of expansions
in the optimizing oscillator basis occurs within a relatively wide range of $%
\omega _{0}$ variations around its already defined values, as we have
previously emphasized. Specifically, practical calculations show that the
value of the basis frequency $\omega _{0}$, calculated according to formula
(47) with parameters (48), provides convergence of the expansions using a
modified oscillator basis of order $N\lesssim 15\div 20$ for  
standard precision computations. Thus, for a fixed value of 
 ${\lambda }$, formula (47) with parameters (48) provides
a very good zeroth approximation for $\omega _{0}$ 
without any further fitting or computations,  
which facilitates practical calculations.
\begin{table}[h] 
\captionsetup{width=0.91\textwidth} 
{\boldmath{ 
\caption{\footnotesize \textbf{Calculated energy values $E_{n}\left( \lambda=1 \right)$  
for the ground state and the ten lowest excited states of the anharmonic 
oscillator at the coupling constant $\lambda=1$. 
$N$ is the order of the applied expansion (38) over the modified 
oscillator basis. The second column presents the optimal values of the 
frequency $\omega _{0}$ of the modified harmonic oscillator basis used for 
the expansion, empirically chosen to ensure the best convergence rate.}}   
}} 
\vspace*{-1.5em} 
 \begin{center} 
{ 
\setlength{\tabcolsep}{25pt} 
{\footnotesize 
\begin{tabular}{@{}lccc@{}}
\hline\hline
$n$ & $\omega _{0}$ & $N$ & $E_{n}\left( \lambda=1 \right)$ \\ \hline
0 & 4.5 & 6  & 0.80377065 \\ 
1 & 4.6 & 5  & 2.73789227 \\ 
2 & 4.9 & 8  & 5.17929169 \\ 
3 & 5.4 & 9  & 7.94240398 \\ 
4 & 5.5 & 10 & 10.96358309 \\ 
5 & 5.8 & 11 & 14.20313910 \\ 
6 & 5.6 & 11 & 17.63404912 \\ 
7 & 5.7 & 11 & 21.23643549 \\ 
8 & 6.0 & 13 & 24.99493641 \\ 
9 & 6.1 & 13 & 28.89725112 \\ 
10 & 6.3 & 15 & 32.93326304 \\ \hline\hline
\end{tabular} }  
\label{tab10} }
 \end{center} 
\end{table}

     As noted earlier, in complete analogy with the above, 
it is also easy to perform the calculation of the excited states based 
on the expansion (38) over the modified oscillator basis, i.e., 
following the main system of algebraic equations (46), with a basis 
frequency $\omega _{0}$ empirically determined. Table 10 presents the results of 
calculations of the eigenvalues $E_{n}\left( \lambda=1 \right)$  for the 
ground state and the ten lowest 
excited states of the anharmonic oscillator for the representative 
coupling constant $\lambda=1$. 
The computation employed standard precision 
to eight significant digits after the decimal for a certain order $N$ of the 
expansion (38) of the system's wave function over the optimizing oscillator 
basis with an empirically determined frequency $\omega _{0}$. 
Here, $N$ is the minimum order 
of the expansion that ensures the required specified accuracy. The calculations 
were performed according to the optimized method discussed above, i.e., 
the energy values were calculated as the eigenvalues of the system of 
equations (46) with the optimized parameter $\omega _{0}$ chosen appropriately.
    
    Table 10 provides clear and illustrative evidence of the remarkably fast 
convergence and excellent efficiency of the method in calculating excited 
states, with an optimally  chosen adjustable frequency parameter $\omega _{0}$, 
as anticipated from the earlier calculations of the ground state values 
of the anharmonic oscillator. In this case, the order $N$ of the optimizing 
oscillator basis required for calculating the energies of the excited states 
needs to be slightly increased compared to that for the ground state. However, 
Table 10 shows that this necessary increase in the order of the modified 
basis used for the calculations is quite moderate. It also demonstrates 
that the optimal values of the empirical frequency parameter $\omega _{0}$  
should be slightly increased to achieve the best rate of 
convergence when calculating excited states, 
compared to the calculation of the ground state. Overall, this increase in the 
parameter $\omega _{0}$ remains fairly small for the chosen 
value of the coupling constant $\lambda=1$. 
At the same time, the study of the general nature of the empirical dependence 
$\omega _{0}\left(n, {\lambda }\right) $   
for different ranges and regimes of the parameters $n$ and ${\lambda }$ 
is a distinct problem requiring further investigation.   

    Nevertheless, a very 
 good zeroth approximation 
for the parameter $\omega _{0}$, at least for not very large values of $n$, would be its 
value obtained from the empirical formulas (47) and (48) for the dependence 
$\omega _{0}\left(0, {\lambda }\right) $    
in the case of the ground state of the system. In addition, as previously noted, 
fixing the empirical parameter $\omega _{0}$ in practical calculations is quite feasible 
with a rather low precision of $\sim 0.2\div 0.7$ or even much lower, which is sufficient 
for reliably determining the region of optimal convergence with respect to $N$   
for both the ground and excited states. Consequently, the proposed 
optimized method generally provides a very convenient, fast, and efficient way to 
calculate the energies of the excited states of the anharmonic oscillator across 
the entire range of variation of the coupling constant. The results of the 
additional calculations of the excited states presented in Table 11 for the 
characteristic value of the coupling constant $\lambda=1000$  from the superstrong coupling 
region confirm this. 
    Moreover, the order $N$ of the modified oscillator basis 
required for the calculations does not increase at all with the growth of the 
parameter ${\lambda }$ and, as before, remains extremely small. Thus, the corresponding 
calculations of the excited states can be effectively performed even in the 
superstrong coupling region using an optimizing basis of extremely small 
dimension, $N\lesssim 20\div 30$. Under these conditions, 
the most economical and truly effective method 
for calculating the low-lying levels is based on selecting the frequency $\omega _{0}$    
according to the empirical formulas (47) and (48), 
without any need for further refinement,  
given that the corresponding expansions already exhibit excellent convergence,   
 and the calculations require only a 
very small number of modified basis functions in any case.
\begin{table}[h] 
\captionsetup{width=0.87\textwidth} 
{\boldmath{ 
\caption{\footnotesize \textbf{Calculated energy values $E_{n}\left( \lambda=1000 \right)$  
for the ground state and the ten lowest excited states of the anharmonic 
oscillator at  $\lambda=1000$. 
$N$ is the order of the applied expansion (38) over the modified 
oscillator basis. The second column lists the optimal values 
of the frequency $\omega _{0}$     
  of the modified harmonic oscillator basis used for 
the expansion, empirically chosen to ensure the best convergence rate.}}   
}} 
\vspace*{-1.5em} 
 \begin{center} 
{ 
\setlength{\tabcolsep}{25pt} 
{\footnotesize 
\begin{tabular}{@{}lccc@{}}
\hline\hline
$n$ & $\omega _{0}$ & $N$ & $E_{n}\left( \lambda=1000 \right)$ \\ \hline
0 & 18.5 & 11  & 6.69422085 \\ 
1 & 19.9 & 11  & 23.97220606 \\ 
2 & 30.2 & 10  & 47.01733873 \\ 
3 & 35.3 & 11  & 73.41911384 \\ 
4 & 36.2 & 12 & 102.51615713 \\ 
5 & 39.3 & 11 & 133.87689122 \\ 
6 & 38.0 & 13 & 167.21225819 \\ 
7 & 41.0 & 12 & 202.31119968 \\ 
8 & 39.7 & 14 & 239.01157755 \\ 
9 & 40.5 & 14 & 277.18416758 \\ 
10 & 39.5 & 16 & 316.72309323 \\ \hline\hline
\end{tabular} }  
\label{tab11} }
 \end{center} 
\end{table}
     
    Obviously, for a specified value of ${\lambda }$ and the already determined
optimal value of the basis frequency $\omega _{0}$ for this case, one can  
 calculate the coefficients $c_{n}$ for a given state, 
which together constitute    
 the system's wave function  in the oscillator
representation corresponding to the modified harmonic oscillator with
frequency $\omega _{0}$, taking into account the parity of the state as well.  
Table 12 lists the coefficients $%
c_{2n}$ from the expansion (38) of the ground state wave function $\psi
_{0}\left( {\lambda ;x}\right) $ of the system in the modified oscillator
basis, calculated in this manner for various values of the parameter $%
{\lambda }$, covering the entire practically significant range of its
variation. 
In these calculations,  
the expansion was performed using 20 modified basis functions.  
\begin{table}[H] 
\captionsetup{width=0.91\textwidth} 
{\boldmath{ 
\caption{\footnotesize \textbf{Values of the coefficients $c_{2n}$,
for the expansion (38) in the modified oscillator basis 
$\left\{ \varphi_{n}\left( \omega _{0};x\right) \right\} $, 
of the ground state wave
function $\psi _{0}\left( {\lambda ;x}\right) $  
of the quartic anharmonic
oscillator for the coupling constant values 
$\lambda =1$, $\lambda =10$, 
$\lambda =100$, $\lambda =1000$, and $\lambda =10000$. 
Frequencies $\omega _{0}$ of the optimized harmonic oscillator basis used for the
expansion were chosen for these cases according to Tables 8 and 9,
respectively equal to 
$\omega _{0}=4.5$, $\omega _{0}=7.5$, 
$\omega_{0}=16.0$, $\omega _{0}=18.5$, and $\omega _{0}=60.5$. The expansion
order in the modified oscillator basis was set to $N=19$, 
corresponding to 20 basis functions.}}  
}}  
\vspace*{-1.5em}
 \begin{center} 
{ 
\setlength{\tabcolsep}{5pt} 
{\footnotesize 
\begin{tabular}{@{}cccccc@{}} 
\hline\hline
$n$ & $c_{2n}\left( \lambda =1\right),$ & $c_{2n}\left( \lambda =10\right),$ 
& $c_{2n}\left( \lambda =100\right),$ & $c_{2n}\left( \lambda
=1000\right),$ & $c_{2n}\left( \lambda =10000\right),$ \\ 
& $\omega _{0}=4.5$ & $\omega _{0}=7.5$ & $\omega _{0}=16.0$ & $\omega
_{0}=18.5$ & $\omega _{0}=60.5$ \\ \hline
0 & 9.55355013e-1 & 9.70655883e-1 & 9.69692134e-1 & 9.99110157e-1 & 
9.84431582e-1 \\ 
1 & 2.84244890e-1 & 2.37539027e-1 & 2.41318709e-1 & 2.02506482e-2 & 
1.75647288e-1 \\ 
2 & 7.89514339e-2 & 3.74117438e-2 & 3.81988355e-2 & -3.60491147e-2 & 
3.57616856e-4 \\ 
3 & 1.62794383e-2 & -8.29521468e-4 & -1.01401691e-3 & 8.17925100e-3 & 
-6.48367316e-3 \\ 
4 & 1.80780224e-3 & -1.42597453e-3 & -1.50315668e-3 & 6.66001463e-4 & 
4.04047971e-4 \\ 
5 & -1.37740275e-4 & 5.90643593e-6 & 1.87534663e-5 & -1.28018030e-3 & 
3.87492026e-4 \\ 
6 & -8.07068788e-5 & 7.93920794e-5 & 8.53217477e-5 & 5.23682996e-4 & 
-1.00195796e-4 \\ 
7 & -3.27100575e-6 & -5.26233008e-6 & -6.86537228e-6 & -7.68673541e-5 & 
-1.30398138e-5 \\ 
8 & 3.33815176e-6 & -5.14988747e-6 & -5.51512709e-6 & -4.54300498e-5 & 
1.33585634e-5 \\ 
9 & 3.32277500e-7 & 1.07229017e-6 & 1.28539912e-6 & 4.15097364e-5 & 
-2.47563602e-6 \\ 
10 & -1.70606806e-7 & 2.65593030e-7 & 2.65016879e-7 & -1.75893435e-5 & 
-7.57963449e-7 \\ 
11 & -1.97791918e-8 & -1.45365311e-7 & -1.68338027e-7 & 3.21697076e-6 & 
5.70625900e-7 \\ 
12 & 1.09204812e-8 & 4.80063336e-9 & 1.01742250e-8 & 1.52593638e-6 & 
-1.14795659e-7 \\ 
13 & 8.77385921e-10 & 1.35307032e-8 & 1.48626978e-8 & -1.79512828e-6 & 
-3.23270182e-8 \\ 
14 & -8.06337466e-10 & -3.89915067e-9 & -4.91627195e-9 & 9.63777679e-7 & 
2.98894155e-8 \\ 
15 & -3.07169601e-12 & -4.35983694e-10 & -3.04869212e-10 & -2.99666349e-7 & 
-8.32306907e-9 \\ 
16 & 6.14995240e-11 & 5.71670561e-10 & 6.62201585e-10 & -3.40100537e-9 & 
-7.72938607e-10 \\ 
17 & -6.40101015e-12 & -1.22635730e-10 & -1.69576799e-10 & 7.59003755e-8 & 
1.67895553e-9 \\ 
18 & -4.24338436e-12 & -3.04678553e-11 & -2.66458376e-11 & -5.66906002e-8 & 
-7.02549038e-10 \\ 
19 & 1.14886008e-12 & 2.36086868e-11 & 2.78333029e-11 & 2.18045343e-8 & 
1.07150382e-10 \\ \hline\hline
\end{tabular} }  
\label{tab12} }
 \end{center} 
\end{table}

    The results in Table 12 confirm the very high convergence rate
of the expansions in terms of the basis $\left\{ \varphi _{n}\left( \omega
_{0};x\right) \right\} _{n=0}^{\infty }$ for any practically relevant  
values of the coupling constant ${\lambda }$, provided that an optimal value
of the frequency $\omega _{0}$ is appropriately chosen. In this case, 
virtually no deterioration   
 in the convergence rate of the
coefficients $c_{2n}$ is observed as ${\lambda }$ increases, 
similar to what is observed when calculating the binding energy.
   Thus, Table 12 provides the tabulated wave function of the system in the
oscillator representation, defined by the Hamiltonian $\mathscr{H}^{\left(
0\right) }\left( \omega _{0}\right) $, for a whole range of typical values
of the coupling constant ${\lambda }$. Any specific value of the wave
function $\psi _{0}\left( {\lambda ;x}\right) $ in the coordinate
representation at a given coordinate $x$ can be easily  
obtained from the expansion (38) for a specified value of the parameter $%
\omega _{0}$. 

     In this context, using 20 basis functions $\varphi _{n}\left( \omega
_{0};x\right) $ in the expansion, the accuracy achieved in calculating
the coefficients $c_{2n}$, i.e., the system's wave function in the optimized
oscillator representation, reaches $\lesssim 10^{-10}\div
10^{-8}$ across the entire range of the coupling constant ${\lambda }$
variation. Table 12 once again clearly demonstrates the excellent quality of
the zeroth approximation used in this approach, that is, in the case of
expansion (38) over the modified oscillator basis with the optimally
chosen value of the parameter $\omega _{0}$. This follows from the fact that
the zeroth expansion coefficient $c_{0}$ contributes in this case at a level
of ${\gtrsim 96\%}$, representing the dominant term, across
all values of the coupling constant ${\lambda }$. Compare this with the
value $c_{0}\left( \lambda =10\right) $ from Table 7, which highlights the clear
trend of the declining quality of the zeroth approximation in the standard
approach, as ${\lambda }$ increases. In this connection, 
it is also interesting and important 
to note that in [67], the asymptotic behavior of the 
coefficients $c_n(\lambda,\omega _{0})$ 
for large $n$ was derived, and their rapid exponential decay with increasing $n$  
was demonstrated, which once again confirms the convergence of 
the main expansion (38) of the wave function of our system in the oscillator basis.
  \begin{figure}[h]
       \centerline{\includegraphics[width=\textwidth]{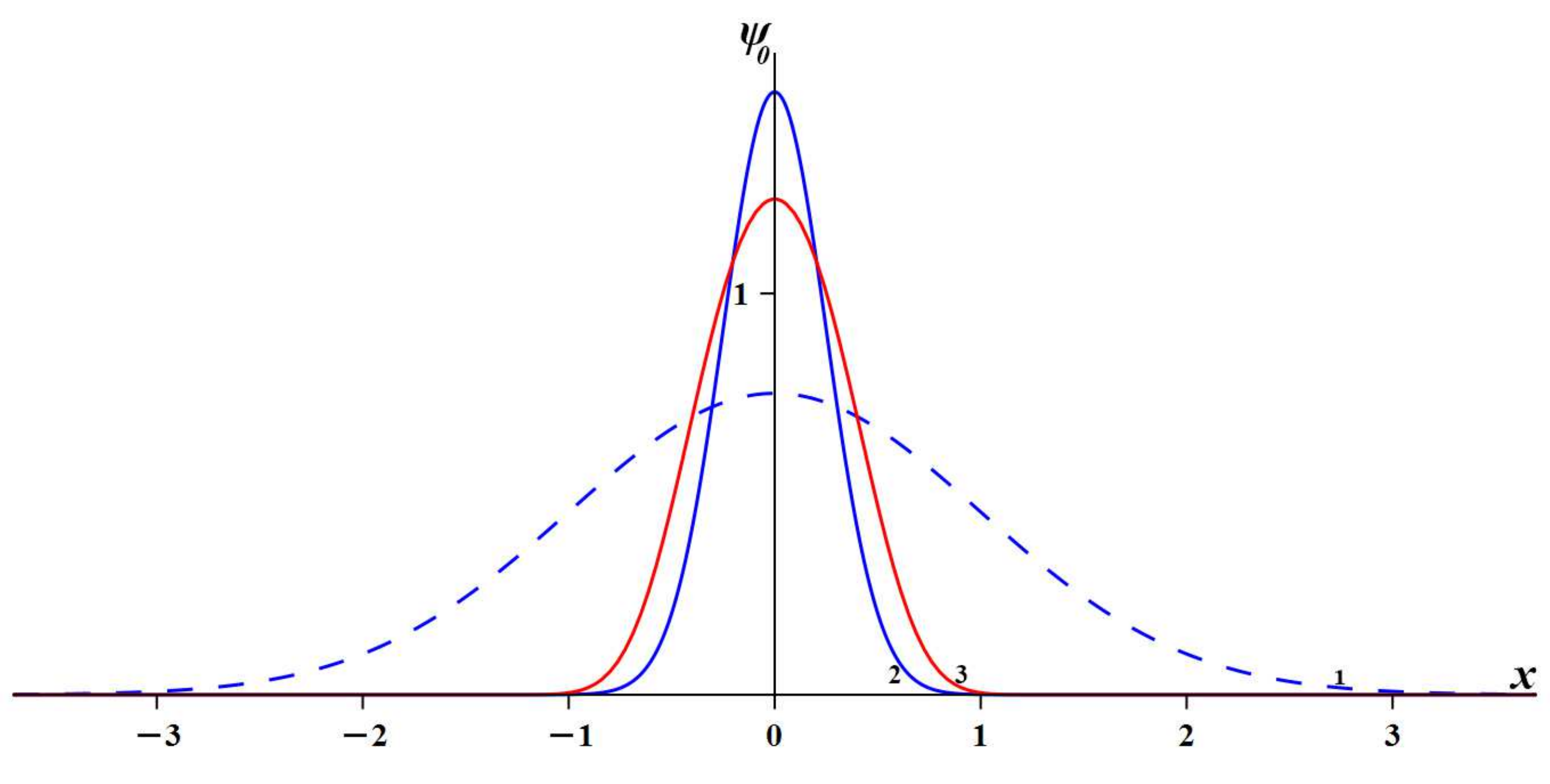}} 
  \caption{The ground state wave functions of the harmonic oscillator, $\varphi
_{0}\left( \omega _{0}=1;x\right) $ at a frequency of $\omega _{0}=1$
(dashed blue curve 1) and $\varphi _{0}\left( \omega _{0}=16;x\right) $ at a
frequency of $\omega _{0}=16$ (solid blue curve 2), along with the ground
state wave function of the quartic anharmonic oscillator $\psi _{0}\left( {%
\lambda =100;x}\right) $ at a coupling constant of ${\lambda =100}$ (red
curve 3). For comparison, see also Fig. 3. \label{f5}}
  \end{figure}

   The same high quality of the zeroth approximation in this approach
is evident from Fig. 5, which presents the system's ground state wave
function in coordinate representation, calculated according to Table 12, for
the coupling constant ${\lambda =100}$, compared with the wave functions of
the harmonic oscillator $\varphi _{0}\left( \omega _{0}=1;x\right) $ and $%
\varphi _{0}\left( \omega _{0}=16;x\right) $. Meanwhile, the last two
functions provide the zeroth approximation respectively in the traditional
approach and in the approach using the expansion over the optimizing
oscillator basis with appropriately chosen 
frequency parameter $\omega _{0}$. 
The superior quality of the zeroth approximation in the second case is
quite obvious from Fig. 5, as also observed in 
Fig. 3, where the corresponding zeroth approximation potentials were
depicted in comparison with the perturbation potential. In fact, Figs. 3 and
5 demonstrate that, in principle, by slight further variation of the
parameter $\omega _{0}$, the zeroth approximation in this case, i.e., for
the coupling constant ${\lambda =100}$, could have been further improved. 
Thus, Table 12 and Fig. 5 reaffirm that the
exceptionally high convergence rate of expansions over the modified
oscillator basis results from the excellent quality of the zeroth
approximation in this approach, achieved by varying the frequency $\omega
_{0}$ of the optimizing oscillator basis $\left\{ \varphi _{n}\left( \omega
_{0};x\right) \right\} _{n=0}^{\infty }$.

   High-precision calculations with many significant digits also
become much more convenient when using the enhanced expansion over the
modified oscillator basis with optimally chosen parameter $\omega _{0}$,
as previously described. Moreover, this approach requires considering
only a relatively small number of basis functions in the expansions. Table
13 presents the results of high-precision calculations, accurate to twenty
significant digits after the decimal, for the ground state energy values $%
E_{0}\left( {\lambda }\right) $ of the anharmonic oscillator, provided for
selected values of ${\lambda }$ across its full 
practical range. The results in Table 13 demonstrate that
high-precision calculations with an accuracy of twenty significant digits
are possible for arbitrarily large values of  ${%
\lambda }$ using a modified oscillator basis of very small dimension $%
N\lesssim 25\div 30$, i.e., approximately with a threefold increase in the
basis dimension compared to cases calculated with the standard accuracy of
eight digits. Also, just as before, it is important that the
required dimension of the oscillator basis for the calculations does not
practically increase with the growth of ${\lambda }$ or 
increases only slightly.
\begin{table}[h] 
\captionsetup{width=0.90\textwidth}
{\boldmath{ 
\caption{\footnotesize \textbf{High-precision calculation results for
the ground state energy $E_{0}$ values of the quartic anharmonic
oscillator with an accuracy of 20 significant digits after the decimal,
given for certain values of the coupling constant $\lambda $. $N$ is
the minimal order of expansion (38) in the modified oscillator basis
required to achieve this degree of accuracy. In the second column, the
optimal values of the frequency $\omega_{0}$ of the modified harmonic
oscillator basis, employed for the expansion, are presented; these values
were chosen empirically to ensure the best convergence rate. Compare with
Table 4.}}  
}} 
\vspace*{-1.5em} 
 \begin{center} 
{ 
\setlength{\tabcolsep}{25pt} 
{\footnotesize 
\begin{tabular}{@{}lccc@{}}
\hline\hline
${\lambda }$ & $\omega _{0}$ & $N$ & $E_{0}$ \\ \hline
0.1 & 2.5 & 14 & 0.55914632718351957672 \\ 
0.25 & 3.7 & 15 & 0.62092702982574866086 \\ 
0.5 & 3.7 & 16 & 0.69617582076514592783 \\ 
1 & 3.7 & 19 & 0.80377065123427376935 \\ 
5 & 7.0 & 19 & 1.22458703605919345913 \\ 
10 & 7.2 & 21 & 1.50497240777889109916 \\ 
50 & 12.0 & 21 & 2.49970877256879391465 \\ 
100 & 13.0 & 26 & 3.13138416493754427316 \\ 
1000 & 22.5 & 29 & 6.69422085050403096950 \\ 
10000 & 69.0 & 22 & 14.39799534352480703004 \\ 
20000 & 87.0 & 22 & 18.13722906686841773519 \\ \hline\hline
\end{tabular} }  
\label{tab13} }
 \end{center} 
\end{table}
     
    Thus, high-precision calculations are quite conveniently feasible within the
framework of the considered approach of expansion in an optimized oscillator
basis, requiring only a small number of basis functions in the 
expansion. Obviously, this is based on the appropriate choice of the 
optimal value of the variable fitting frequency $\omega _{0}$, just as in
the past. Notably, even at very
small values of the coupling constant ${\lambda }$, it is highly beneficial 
and effective to choose values of the frequency parameter $\omega _{0}$
greater than one, as is clear from Table 13. It is worth mentioning that 
there is some variability in the values of the parameter $\omega _{0}$
depending on the required accuracy of the calculations. However, this is not
a problem due to the relatively low sensitivity of the convergence rate of
the expansions to the parameter $\omega _{0}$, and due to the very rapid
convergence of the expansions at values close to the optimal $\omega _{0}$,
ensuring fast calculations. 

     In recent studies [41, 42], a specific benchmark 
high-precision value for the ground state
energy $E_{0}\left( {\lambda =1/4}\right) $ of the quartic anharmonic
oscillator was calculated at a coupling constant value of ${\lambda =1/4}$.
In those studies, this value was determined with an accuracy of 40 significant digits
in [41] and with an accuracy of 246 significant digits in  
[42]. For validation and demonstration, we calculated this
value according to the above-described method of expansion in an optimized
oscillator basis with an accuracy of 250 significant digits: 
\begin{equation}
\begin{aligned} E_{0}\left( { \frac{1}{4} }\right)
&=0.620927029825748660858035732987120698200017253619138982542367 \\
&3250629627481887688839793913513034794560836016187600734766248910 \\
&8576830809906593840258008453039702473747434766340695449307556609 \\
&30523968593024724863926019751363572931088715294391170922759124~.
\end{aligned}  \tag{49}
\end{equation}%
Our result in (49) exactly reproduced the first 245 significant digits from 
[42], and the one-unit difference in the last significant digit of  
[42] is evidently due to the insufficient precision there in determining
that final digit. 

    As an illustration, Table 14 presents the required order $N$ 
of the optimizing oscillator basis needed to achieve a specified
 precision level, measured in \textquotedblleft
Digits\textquotedblright\ of significant digits, when computing the energy
value $E_{0}\left( {\lambda =1/4}\right) $. In this case, the value of the
fitting parameter $\omega _{0}$ was fixed at $\omega _{0}=3.7$  
as per Table 13. 
For comparison, Table 14 also lists the number of Lagrange mesh
points $Y$ required to achieve the same precision 
 in the Lagrange mesh method, according to  
[42]. The results in Table 14 show that, overall, the increase in the
number of basis functions in the modified expansion (38), required to
achieve higher precision, is quite moderate. 
Thus, the results in Tables 13 and 14, 
along with the comparison of the calculated energy value $%
E_{0}\left( {\lambda =1/4}\right) $ (49) with its benchmark 
reference value [41, 42], demonstrate the convenient 
feasibility  of performing  high-precision calculations 
for the energy levels of the anharmonic
oscillator using the expansion method (38) 
in the modified oscillator basis. 
Naturally, the optimal fitting frequency $\omega _{0}$ 
of this basis must be chosen each time, which can be quickly accomplished by
conducting  preliminary  calculations with a small number 
of significant digits or simply by employing the empirical formulas 
(47) and (48) to fix $\omega _{0}$. 
Moreover, a highly 
convenient feature of this method is its cost-effective 
applicability  at any, arbitrarily large, values of the coupling constant ${%
\lambda }$, practically without increasing the basis dimension $N$, thanks
to the optimal choice of the parameter $\omega _{0}$.  
\begin{table}[h] 
\captionsetup{width=0.87\textwidth} 
{\boldmath{ 
\caption{\footnotesize \textbf{The required minimum order $N$ of
the expansion (38) in the modified oscillator basis (with the chosen
basis frequency $\omega_{0}=3.7$) and the necessary number $Y$ of
Lagrange mesh points, according to [42], to achieve the specified
precision in Digits of significant digits in the high-precision calculation
of the ground state energy $E_{0}\left( {\lambda =1/4}\right) $ of the
quartic anharmonic oscillator at the coupling constant $\lambda =1/4$.}}  
}} 
\vspace*{-1.5em} 
 \begin{center} 
{ 
\setlength{\tabcolsep}{30pt}
{\footnotesize 
\begin{tabular}{@{}ccc@{}}
\hline\hline
{Digits} & $N$ & $Y$ \\ \hline
8 & 7 & 25 \\ 
14 & 12 & 50 \\ 
20 & 15 & 75 \\ 
25 & 21 & 100 \\ 
33 & 27 & 150 \\ 
44 & 38 & 200 \\ 
50 & 43 & 250 \\ 
59 & 52 & 300 \\ 
72 & 66 & 400 \\ 
87 & 82 & 500 \\ 
145 & 152 & 1000 \\ 
237 & 284 & 1900 \\ 
246 & 297 & 2000 \\ 
250 & 304 & --- \\ \hline\hline
\end{tabular} }  
\label{tab14} }
 \end{center} 
\end{table}

{
\begin{center}
{\textbf{ 7. Discussion and Conclusions }}  
\end{center}
{\noindent
Overall, it should be noted that a great variety of diverse methods have
been proposed for studying the properties and calculating the
characteristics of the quantum anharmonic oscillator. These methods
differ from each other both in the complexity of their practical application
and in their effectiveness. At the same time, each of the numerous proposed methods
has both its advantages and disadvantages. It should also be emphasized that
a number of these methods can work and be effectively applied only in the
region of relatively weak coupling. Additionally, it is very significant
that many of the methods are capable of being effectively used only for
calculating the energy values of the anharmonic oscillator states, but not
for calculating the wave functions of these states. From our standpoint, all these
various methods and approaches to studying the anharmonic oscillator should
not be seen as competitors, but rather as conveniently complementing one
another --- especially since the development and application of any of them
may prove to be relevant and useful for studying other similar systems in
the future. Among the recently developed and applied methods for calculating
the properties of the anharmonic oscillator, one can note such modern highly
efficient computational methods as the Lagrange mesh method [10, 42],
the Riccati-Pad\'{e} method [36, 70], and the double exponential
Sinc-collocation method [9].} }

Special mention should be made of different implementations of 
the Rayleigh-Ritz (RR) variational method in the study of the anharmonic 
oscillator [66--69], as a number of these schemes included the use of 
an oscillator basis with its frequency optimized in various ways, as noted above. 
More detailed information on this subject, along with further references and 
discussion, can be found in [68, 69]. It should be noted, however, that the basis 
expansion method used in our work is generally not equivalent to the RR method, 
even though the corresponding truncated systems of algebraic equations 
formally coincide. Overall, our method is more general and rigorous than 
the RR method, providing several advantages and additional benefits. 
A more detailed and comprehensive comparative analysis will be presented in 
our future works. Here, we simply note that the very existence of nearly 
a dozen different approaches to optimizing the oscillator basis frequency 
within the RR scheme highlights the incompleteness and unresolved nature 
of this issue in the RR method.  
   
In general, it is important to stress that the method employed in our study, 
involving the convergent expansion of the system's wave function over 
the oscillator basis, is a well-established approach to solving 
quantum problems, which is also one of the simplest in its application. 
The numerous successful  
uses of this method [47--51] demonstrate that the
oscillator-basis expansion approach offers a range of advantages in practice 
and is exceptionally convenient and appealing due to its simplicity,
versatility, and generality.  
Also, one of the highly significant advantages of this method is its  
ability to perform simple and reliable calculations not only of  
the energy values of the anharmonic oscillator but also of 
 the wave functions corresponding to quantum states  
of the oscillator, a capability that many other approaches lack. 
Thus, in  addition to other benefits, 
the presented method offers a very convenient and accurate way to calculate
the system's wave function for any value of the coupling constant ${\lambda }$. 
 
   Altogether, the discussed oscillator-basis expansion method enables  
relatively simple calculation of the energy values of the ground and 
excited states of the quartic anharmonic oscillator across a wide range of
the coupling constant ${\lambda }$ variation. The foundation of the method,
in this case, lies in the diagonalization of a finite symmetric matrix of
the system's Hamiltonian in its oscillator representation. 
In other words, in this approach, 
determining the eigenenergy levels and eigenwave
functions of the anharmonic oscillator  reduces to
 finding the eigenvalues and eigenvectors of a finite real
symmetric square matrix of the system's Hamiltonian in the harmonic
oscillator representation. Moreover, the problem of
obtaining the eigenvalues and eigenvectors of a certain finite-order $N$
square real symmetric matrix is a classic problem, 
with well-developed methods for its solution. 
      The convergence rate of oscillator expansions
in specific calculations proves to be very high across a wide range of
variations in the oscillator coupling constant ${\lambda }$, although it
naturally decreases as ${\lambda }$ increases. Generally, for calculations
of the energy values of low-lying levels with standard precision to eight
significant digits across the whole practically necessary range of the
parameter  ${\lambda }$, choosing an oscillator expansion order of $%
N\lesssim 1000$ is sufficient. Thus, the results of our
calculations demonstrate the capability of the considered method to reliably
obtain energy values for a wide spectrum of excited oscillator states over a
broad range of ${\lambda }$ using a basis of oscillator
functions of moderate dimension $N\lesssim 1000$. 
Overall, the presented method for studying the anharmonic oscillator, 
based on the convergent expansion of the system's wave function 
in the oscillator basis, provides a convenient way to calculate 
all physical characteristics of the system, including wave functions 
for various states, using relatively low-order expansions.
Furthermore, due to the noted excellent computational development of methods
for solving the problem of determining eigenvalues and eigenvectors of
matrices, the specific numerical calculation of the system's energy
eigenvalues and the coefficients of the oscillator expansion, expressing the
system's wave function in the oscillator representation, does not pose any
 difficulties.

Our main central result  is the method we propose for
significantly accelerating the convergence of oscillator expansions. This is
achieved by varying the frequency parameter $\omega_0$ of the modified oscillator
basis functions with arbitrary frequency, which we suggest using for the
expansion. Since the initial consideration corresponded to 
the limiting particular case of $\omega_0=\omega=1$, 
the proposed modification of the approach is a natural generalization 
of the original formalism. This modified approach allows  
a substantial increase in the efficiency and performance 
 of the applied method through a more effective and
rational choice of the frequency parameter $\omega_0$ of the zeroth-approximation 
Hamiltonian. Specifically, through a reasonable choice of the frequency of
the zeroth-approximation Hamiltonian, one can achieve a much better and more
suitable zeroth approximation, bringing it significantly closer to the original total
Hamiltonian. Thus, a more effective choice of the frequency parameter of the
zeroth approximation, namely, increasing it for larger fixed values of the
coupling constant ${\lambda }$, can significantly enhance the quality and
adequacy of the zeroth approximation. The obtained numerical calculation
results demonstrate an exceptionally high convergence rate and excellent
efficiency of the proposed approach, with an appropriate choice of the
variable frequency parameter for a given specified value of 
 ${\lambda }$. It turns out that the convergence rate, achieved by
varying and optimally choosing the basis frequency for the expansion, does
not decrease at all and remains consistently high across the
entire range of variation of ${\lambda }$. 
      
      As a result, the
proposed method for accelerating convergence allows for the calculation of
the anharmonic oscillator's energy values with standard precision throughout
the entire range of variation of the coupling
constant ${\lambda }$, by using only a very small number of basis functions 
$N\lesssim 10$,  
which, notably, does not increase as ${\lambda }$ increases.
In addition, the convergence of the system's wave function in 
the oscillator representation is also significantly accelerated, 
 enabling reliable and precise calculations of 
the oscillator wave functions for all  
 values of ${\lambda }$.
Furthermore, in this enhanced optimizing approach, 
high-precision calculations involving a large
number of significant digits become much more convenient, as they
use only a relatively small number of basis functions in the expansions.  
Thus, practical calculations for the entire range of
 ${\lambda }$ become 
feasible based on the proposed approach of varying the frequency of the
optimizing oscillator basis, by utilizing a basis of very small dimension.

Overall, the approach based on the expansion over an optimizing
oscillator basis with a variable fitting frequency actually solves the
problem of strong coupling in this case, as the number of basis functions
required for calculations does not increase with the growth of the ${\lambda 
}$ parameter. Consequently, this modified approach provides an essentially 
complete, quite simple, and efficient solution to the problem of the quartic
anharmonic oscillator, enabling easy computation of all its
physical properties, including the energies of the ground and excited
states, as well as the wave functions of these states, for arbitrary 
 values of the coupling constant. On the whole, the classical method of 
solving quantum mechanical problems based on the convergent 
 expansion of the system's wave function 
over the oscillator basis has once again demonstrated its convenience,
simplicity, and extremely high efficiency in the case considered. Moreover,
its modification involving the use of an expansion over a modified
oscillator basis with 
a suitably chosen 
 variable fitting frequency proved to be
especially reliable and fruitful. This enhanced expansion method can
undoubtedly be successfully applied to solve many other similar problems as
well, particularly for the study and calculation of characteristics of other
oscillator models with more singular perturbations. 

In conclusion, we would like to express our gratitude to V. Vasilevsky for valuable 
discussions of the results. The present study received partial support from the 
Fundamental Research Program of the Physics and Astronomy Division of 
the National Academy of Sciences of Ukraine 
(projects No. 0117U000239 and No. 0122U000886) and the Simons Foundation.

\begin{center}
{\normalsize REFERENCES}
\end{center}
\begin{enumerate} 
\item C. M. Bender and T. T. Wu, \textit{Phys. Rev.} \textbf{184},
1231 (1969).

\item F. T. Hioe, D. MacMillen and E. W. Montroll, \textit{Phys. Rep.} 
\textbf{43}, 305 (1978). 

\item D. I. Kazakov and D. V. Shirkov, \textit{Fortschr. Phys.} 
\textbf{28}, 465 (1980).

\item B. Simon, \textit{Int. J. Quantum Chem.} \textbf{21}, 3 (1982).

\item C. Itzykson and J.-B. Zuber, \textit{Quantum Field Theory}
(McGraw-Hill, New York, 1980). 

\item G. A. Arteca, F. M. Fern\'{a}ndez and E. A. Castro, \textit{%
Large Order Perturbation Theory and Summation Methods in Quantum Mechanics}
(Springer-Verlag, Berlin, 1990). 

\item J. C. Le Guillou and J. Zinn-Justin (eds.), \textit{Large-Order
Behaviour of Perturbation Theory} (Elsevier Science Publishers, Amsterdam,
1990). 

\item B. Simon, \textit{Bull. Am. Math. Soc.} \textbf{24}, 303 (1991).

\item P. J. Gaudreau, R. M. Slevinsky and H. Safouhi, 
\textit{Ann. Phys.} \textbf{360}, 520 (2015). 

\item A. V. Turbiner and J. C. del Valle Rosales, 
\textit{Quantum Anharmonic Oscillator} (World Scientific, Singapore, 2023). 

\item G. L\'{e}vai and J. M. Arias, \textit{Phys. Rev. C} \textbf{81}, 044304 (2010). 

\item A. A. Raduta and P. Buganu, \textit{J. Phys. G} \textbf{40}, 025108 (2013).

\item R. Budaca, \textit{Eur. Phys. J. A} \textbf{50}, 87 (2014). 

\item P. Buganu and R. Budaca, \textit{Phys. Rev. C} \textbf{91}, 014306 (2015). 

\item M. M. Hammad, M. M. Yahia, H. A. Motaweh and S. B. Doma, 
\textit{Nucl. Phys. A} \textbf{1004}, 122036 (2020).  

\item J. J. Loeffel, A. Martin, B. Simon and A. S. Wightman, 
\textit{Phys. Lett. B} \textbf{30}, 656 (1969). 

\item B. Simon, \textit{Ann. Phys.} \textbf{58}, 76 (1970). 

\item C. M. Bender and T. T. Wu, \textit{Phys. Rev. Lett.} 
\textbf{27}, 461 (1971). 

\item C. M. Bender and T. T. Wu, \textit{Phys. Rev. D} \textbf{7},
1620 (1973). 

\item S. N. Biswas, K. Datta, R. P. Saxena, P. K. Srivastava and V. S. Varma, 
\textit{J. Math. Phys.} \textbf{14}, 1190 (1973). 

\item F. T. Hioe and E. W. Montroll, \textit{J. Math. Phys.} 
\textbf{16}, 1945 (1975). 

\item K. Banerjee, \textit{Lett. Math. Phys.} \textbf{1}, 323 (1976). 

\item A. V. Turbiner and A. G. Ushveridze, \textit{J. Math. Phys.} 
\textbf{29}, 2053 (1988). 

\item F. Vinette and J. \v{C}\'{\i}\v{z}ek, 
\textit{J. Math. Phys.} \textbf{32}, 3392 (1991). 

\item E. J. Weniger, J. \v{C}\'{\i}\v{z}ek and F. Vinette, 
\textit{Phys. Lett. A} \textbf{156}, 169 (1991).  

\item E. J. Weniger, J. \v{C}\'{\i}\v{z}ek and F. Vinette, 
\textit{J. Math. Phys.} \textbf{34}, 571 (1993). 

\item F. M. Fern\'{a}ndez and R. Guardiola, \textit{J. Phys. A} 
\textbf{26}, 7169 (1993). 

\item E. J. Weniger, \textit{Ann. Phys.} 
\textbf{246}, 133 (1996). 

\item F. M. Fern\'{a}ndez and R. Guardiola, \textit{J. Phys. A} 
\textbf{30}, 7187 (1997). 

\item A. V. Turbiner, \textit{Lett. Math. Phys.} \textbf{74}, 169 (2005). 

\item A. Banerjee, \textit{Mod. Phys. Lett. A} \textbf{20}, 3013
(2005).  
 
\item E. Z. Liverts, V. B. Mandelzweig and F. Tabakin, \textit{J.
Math. Phys.} \textbf{47}, 062109 (2006). 

\item A. J. Sous, \textit{Mod. Phys. Lett. A} \textbf{21}, 1675
(2006). 

\item H. Ciftci, \textit{Mod. Phys. Lett. A} \textbf{23}, 261 (2008). 

\item H. Ezawa, M. Saito and T. Nakamura, \textit{J. Phys. Soc. Japan}
\textbf{83}, 034003 (2014). 

\item F. M. Fern\'{a}ndez and J. Garcia, \textit{Acta Polytech.} 
\textbf{57}, 391 (2017). 

\item T. Sulejmanpasic and M. \"{U}nsal, \textit{Comp. Phys. Com.} 
\textbf{228}, 273 (2018).   

\item H. Mutuk, \textit{Mod. Phys. Lett. A} \textbf{34}, 1950088
(2019).

\item J. C. del Valle and A. V. Turbiner, 
\textit{Int. J. Mod. Phys. A} \textbf{34}, 1950143 (2019).

\item J. C. del Valle and A. V. Turbiner, 
\textit{Int. J. Mod. Phys. A} \textbf{35}, 2050005 (2020). 

\item P. Okun and K. Burke, \textit{Int. J. Quantum Chem.} 
\textbf{121}, e26554 (2021).  

\item A. V. Turbiner and J. C. del Valle, \textit{Int. J. Quantum
Chem.} \textbf{121}, e26766 (2021).

\item A. V. Turbiner and J. C. del Valle, 
\textit{J. Phys. A} \textbf{54}, 295204 (2021).  

\item A. V. Turbiner and J. C. del Valle, \textit{Acta Polytech.} 
\textbf{62}, 208 (2022). 

\item V. A. Babenko and N. M. Petrov, \textit{Nucl. Phys. At. Energy} 
\textbf{22}, 127 (2021). 

\item V. A. Babenko and N. M. Petrov, \textit{Mod. Phys. Lett. A} 
\textbf{37}, 2250172 (2022). 

\item V. Vasilevsky, A. V. Nesterov, F. Arickx and J. Broeckhove, 
\textit{Phys. Rev. C} \textbf{63}, 034606 (2001). 

\item V. Vasilevsky, A. V. Nesterov, F. Arickx and J. Broeckhove, 
\textit{Phys. Rev. C} \textbf{63}, 034607 (2001). 

\item F. Arickx, J. Broeckhove, A. Nesterov, V. Vasilevsky and W. Vanroose, 
The Modified J-Matrix Approach for Cluster Descriptions of Light Nuclei, 
in \textit{The J-Matrix Method}, 
eds. A. D. Alhaidari et al. (Springer, Berlin, 2008) p. 269. 

\item A. V. Nesterov, F. Arickx, J. Broeckhove and V. Vasilevsky, 
\textit{Phys. Part. Nucl.} \textbf{41}, 716 (2010). 

\item M. Moshinsky and Y. F. Smirnov, 
\textit{The Harmonic Oscillator in Modern Physics} 
(Harwood Academic Publishers, Amsterdam, 1996).  

\item N. W. Bazley and D. W. Fox, \textit{Phys. Rev.} \textbf{124},
483 (1961). 

\item S. I. Chan, D. Stelman and L. E. Thompson, \textit{J. Chem. Phys.} 
\textbf{41}, 2828 (1964). 

\item P.-F. Hsieh and Y. Sibuya, 
\textit{J. Math. Anal. Appl.} \textbf{16}, 84 (1966).  

\item F. J. Dyson, \textit{Phys. Rev.} \textbf{85}, 631 (1952). 

\item A. V. Turbiner, \textit{Sov. Phys. Usp.} \textbf{27}, 668
(1984). 

\item A. S. Davydov, 
\textit{Quantum Mechanics} (Pergamon Press, Oxford, 1976). 

\item I. O. Vakarchuk, 
\textit{Quantum Mechanics} (Ivan Franko National University, Lviv, 2012). 

\item R. McWeeny and C. A. Coulson, 
\textit{Math. Proc. Cambridge} \textbf{44}, 413 (1948). 

\item E. B. Wilson, Jr., J. C. Decius and P. C. Cross,  
\textit{Molecular Vibrations} (McGraw-Hill, New York, 1955).  

\item A. H. Nayfeh, 
\textit{Introduction to Perturbation Techniques} (Wiley, New York, 1981). 

\item F. Verhulst, 
\textit{Nonlinear Differential Equations and Dynamical Systems} (Springer, Berlin, 1996). 

\item V. S. Vasilevsky and F. Arickx, 
\textit{Phys. Rev. A} \textbf{55}, 265 (1997). 

\item A. D. Alhaidari, E. J. Heller, H. A. Yamani and M. S. Abdelmonem (eds.), 
\textit{The J-Matrix Method} (Springer, Berlin, 2008).  

\item V. S. Vasilevsky and M. D. Soloha-Klymchak, \textit{Ukr. J. Phys.} 
\textbf{60}, 297 (2015). 

\item R. Balsa, M. Plo, J. G. Esteve and A. F. Pacheco, 
\textit{Phys. Rev. D} \textbf{28}, 1945 (1983). 

\item R. N. Hill, 
\textit{Phys. Rev. A} \textbf{51}, 4433 (1995). 

\item P. Ko\'{s}cik and A. Okopi\'{n}ska, 
\textit{J. Phys. A} \textbf{40}, 10851 (2007).

\item W.-F. Lu, C. K. Kim and K. Nahm, 
\textit{J. Phys. A} \textbf{40}, 14457 (2007).

\item F. M. Fern\'{a}ndez, Q. Ma and R. H. Tipping, \textit{Phys. Rev. A} 
\textbf{40}, 6149 (1989). 
\end{enumerate}        
            
\end{document}